\documentclass[]{aastex631}

\usepackage[utf8]{inputenc}
\usepackage{natbib}
\usepackage{xcolor}
\usepackage{hyperref}
\usepackage{amsmath}
\usepackage{graphicx}
\usepackage[caption=false]{subfig}
\usepackage[utf8]{inputenc}

\usepackage[mathscr]{euscript}

\newcommand{\ebno}[0]{4584~} 

\newcommand{\tess}[0]{\textsl{TESS}~}
\newcommand{\tessnts}[0]{\textsl{TESS}}
\newcommand{\kepler}[0]{\textsl{Kepler}~}
\newcommand{\keplernts}[0]{\textsl{Kepler}}
\newcommand{\twog}[0]{\textit{two-Gaussian} }
\newcommand{\twogs}[0]{\textit{two-Gaussians} }
\newcommand{\twognts}[0]{\textit{two-Gaussian}}
\newcommand{\polyfit}[0]{\textit{polyfit} }
\newcommand{\polyfitnts}[0]{\textit{polyfit}}
\newcommand{\rev}[1]{#1}

\received{}
\revised{}
\accepted{}
\submitjournal{ApJS}

\shorttitle{TESS Eclipsing Binary Stars}
\shortauthors{Prša et al.}

\begin{document}

\title{TESS Eclipsing Binary Stars. I. Short cadence observations of \ebno eclipsing binaries in Sectors 1--26}

\correspondingauthor{Andrej Prša}
\email{aprsa@villanova.edu}


\author[0000-0002-1913-0281]{Andrej Pr\v sa}
\affiliation{Villanova University, Dept.~of Astrophysics and Planetary Sciences, 800 E.\ Lancaster Ave, Villanova, PA 19085, USA}

\author[0000-0002-9739-8371]{Angela Kochoska}
\affiliation{Villanova University, Dept.~of Astrophysics and Planetary Sciences, 800 E.\ Lancaster Ave, Villanova, PA 19085, USA}

\author[0000-0002-5442-8550]{Kyle E.~Conroy}
\affiliation{Villanova University, Dept.~of Astrophysics and Planetary Sciences, 800 E.\ Lancaster Ave, Villanova, PA 19085, USA}

\author[0000-0002-9138-9028]{Nora Eisner}
\affiliation{University of Oxford, Department of Physics, Keble Rd, Oxford OX1 3RH, UK}

\author[0000-0003-3244-5357]{Daniel R. Hey}
\affiliation{School of Physics, Sydney Institute for Astronomy (SIfA), The University of Sydney, NSW 2006, Australia}

\author[0000-0002-9241-3894]{Luc IJspeert}
\affiliation{KU Leuven, Institute of Astronomy, Celestijnenlaan 200D, 3001, Leuven, Belgium}

\author[0000-0002-0493-1342]{Ethan Kruse}
\affiliation{NASA Goddard Space Flight Center, 8800 Greenbelt Road, Greenbelt, MD 20771, USA}


\author[0000-0003-0556-027X]{Scott W. Fleming}
\affiliation{Space Telescope Science Institute, 3700 San Martin Dr, Baltimore, MD 21218, USA}

\author[0000-0002-3054-4135]{Cole Johnston}
\affiliation{Radboud University Nijmegen, Department of Astrophysics, IMAPP, P.O.~Box 9010, 6500 GL Nijmegen, the Netherlands}
\affiliation{KU Leuven, Institute of Astronomy, Celestijnenlaan 200D, 3001, Leuven, Belgium}

\author[0000-0002-2607-138X]{Martti H.~Kristiansen}
\affiliation{Brorfelde Observatory, Observator Gyldenkernes Vej 7, DK-4340 Tølløse, Denmark}
\affiliation{DTU Space, National Space Institute, Technical University of Denmark, Elektrovej 327, DK-2800 Lyngby, Denmark}

\author[0000-0002-8527-2114]{Daryll LaCourse}
\affiliation{Amateur Astronomer, 7507 52nd Pl NE, Marysville, WA 98270}

\author[0000-0002-4113-2654]{Danielle Mortensen}
\affiliation{Villanova University, Dept.~of Astrophysics and Planetary Sciences, 800 E.\ Lancaster Ave, Villanova, PA 19085, USA}

\author[0000-0002-3827-8417]{Joshua Pepper}
\affiliation{Lehigh University, Department of Physics, 13 Memorial Drive East, Bethlehem, PA 18015}

\author[0000-0002-3481-9052]{Keivan G.\ Stassun}
\affiliation{Vanderbilt University, Department of Physics \& Astronomy, 6301 Stevenson Center Ln., Nashville, TN 37235, USA}

\author[0000-0002-5286-0251]{Guillermo Torres}
\affiliation{Center for Astrophysics $\vert$ Harvard \& Smithsonian, 60 Garden Street, Cambridge, MA 02138, USA}


\author[0000-0001-6566-7568]{Michael Abdul-Masih}
\affiliation{European Southern Observatory, Alonso de Cordova 3107, Vitacura, Casilla 19001, Santiago de Chile, Chile}

\author[0000-0002-0568-6000]{Joheen Chakraborty}
\affiliation{Columbia University, Department of Astronomy, 550 W 120th Street, New York NY 10027}

\author[0000-0002-5665-1879]{Robert Gagliano}
\affiliation{Amateur Astronomer, Glendale, Arizona}

\author[0000-0002-0951-2171]{Zhao Guo}
\affiliation{University of Cambridge, Department of Applied Mathematics and Theoretical Physics, Centre for Mathematical Sciences, Wilberforce Road, Cambridge CB3 0WA, UK}

\author[0000-0001-5473-856X]{Kelly Hambleton}
\affiliation{Villanova University, Dept.~of Astrophysics and Planetary Sciences, 800 E.\ Lancaster Ave, Villanova, PA 19085, USA}

\author[0000-0002-8692-2588]{Kyeongsoo Hong}
\affiliation{Institute for Astrophysics, Chungbuk National University, Chungdae-ro 1, Seowon-Gu, Cheongju 28644, Republic of Korea}

\author[0000-0003-3988-3245]{Thomas Jacobs}
\affiliation{Amateur Astronomer, 12812 SE 69th Place, Bellevue, WA 98006}

\author[0000-0003-3947-5946]{David Jones}
\affiliation{Instituto de Astrofísica de Canarias, E-38205 La Laguna, Tenerife, Spain}
\affiliation{Departamento de Astrofísica, Universidad de La Laguna, E-38206 La Laguna, Tenerife, Spain}

\author[0000-0001-9786-1031]{Veselin Kostov}
\affiliation{NASA Goddard Space Flight Center, 8800 Greenbelt Road, Greenbelt, MD 20771, USA}
\affiliation{SETI Institute}

\author[0000-0002-5739-9804]{Jae Woo Lee}
\affiliation{Korea Astronomy and Space Science Institute, Optical Division, 776 Daedeok-daero, Yuseong-gu, Daejeon 34055, Republic of Korea}

\author{Mark Omohundro}
\affiliation{Citizen scientist, c/o Zooniverse, Department of Physics, University of Oxford, Denys Wilkinson Building, Keble Road, Oxford, OX1 3RH, UK}

\author[0000-0001-9647-2886]{Jerome A.~Orosz}
\affiliation{San Diego State University, Department of Astronomy, 5500 Campanile Dr., San Diego, CA 92182-1221, USA}

\author[0000-0002-3221-3874]{Emma J. Page}
\affiliation{Lehigh University, Department of Physics, 13 Memorial Drive East, Bethlehem, PA 18015}

\author[0000-0003-0501-2636]{Brian P. Powell}
\affiliation{NASA Goddard Space Flight Center, 8800 Greenbelt Road, Greenbelt, MD 20771, USA}

\author[0000-0003-3182-5569]{Saul Rappaport}
\affiliation{Department of Physics and Kavli Institute for Astrophysics and Space Research, Massachusetts Institute of Technology, Cambridge, MA 02139, USA}

\author[0000-0002-5005-1215]{Phill Reed}
\affiliation{Department of Physical Sciences, Kutztown University, Kutztown, PA, 19530, USA}

\author[0000-0002-2942-8399]{Jeremy Schnittman}
\affiliation{NASA Goddard Space Flight Center, Astrophysics Science Division, 8800 Greenbelt Rd, Greenbelt, MD 20771, USA}

\author[0000-0002-1637-2189]{Hans Martin Schwengeler}
\affiliation{Citizen Scientist, Zehntenfreistrasse 11, CH-4103 Bottmingen, Switzerland}

\author[0000-0002-1836-3120]{Avi Shporer}
\affiliation{Massachusetts Institute of Technology, Department of Physics and Kavli Institute for Astrophysics and Space Research, Cambridge, MA 02139, USA}

\author[0000-0002-0654-4442]{Ivan A.~Terentev}
\affiliation{Citizen Scientist, Moskovskaya 8, 185031 Petrozavodsk, Russia}

\author[0000-0001-7246-5438]{Andrew Vanderburg}
\affiliation{Department of Physics and Kavli Institute for Astrophysics and Space Research, Massachusetts Institute of Technology, 77 Massachusetts Avenue, Cambridge, MA 02139, USA}

\author[0000-0003-2381-5301]{William F.~Welsh}
\affiliation{San Diego State University, Department of Astronomy, 5500 Campanile Dr., San Diego, CA 92182-1221, USA}


\author[0000-0003-1963-9616]{Douglas A.~Caldwell}
\affiliation{SETI Institute, 189 Bernardo Ave, Suite 200, Mountain View, CA 94043, USA}
\affiliation{NASA Ames Research Center, Moffett Field, CA 94035, USA}

\author{John~P.~Doty}
\affiliation{Noqsi Aerospace Ltd., 15 Blanchard Avenue, Billerica, MA 01821, USA}

\author[0000-0002-4715-9460]{Jon~M.~Jenkins}
\affiliation{NASA Ames Research Center, Moffett Field, CA 94035, USA}

\author[0000-0001-9911-7388]{David~W.~Latham}
\affiliation{Harvard-Smithsonian Center for Astrophysics, 60 Garden St, Cambridge, MA 02138, USA}

\author[0000-0003-2058-6662]{George~R.~Ricker}
\affiliation{Department of Physics and Kavli Institute for Astrophysics and Space Research, Massachusetts Institute of Technology, Cambridge, MA 02139, USA}

\author[0000-0002-6892-6948]{Sara~Seager}
\affiliation{Department of Physics and Kavli Institute for Astrophysics and Space Research, Massachusetts Institute of Technology, Cambridge, MA 02139, USA}
\affiliation{Department of Earth, Atmospheric and Planetary Sciences, Massachusetts Institute of Technology, Cambridge, MA 02139, USA}
\affiliation{Department of Aeronautics and Astronautics, MIT, 77 Massachusetts Avenue, Cambridge, MA 02139, USA}

\author{Joshua~E.~Schlieder}
\affiliation{NASA Goddard Space Flight Center, 8800 Greenbelt Rd, Greenbelt, MD 20771, USA}

\author{Bernie~Shiao}
\affiliation{Space Telescope Science Institute, 3700 San Martin Drive, Baltimore, MD, 21218, USA}

\author[0000-0001-6763-6562]{Roland~Vanderspek}
\affiliation{Department of Physics and Kavli Institute for Astrophysics and Space Research, Massachusetts Institute of Technology, Cambridge, MA 02139, USA}

\author[0000-0002-4265-047X]{Joshua~N.~Winn}
\affiliation{Department of Astrophysical Sciences, Princeton University, 4 Ivy Lane, Princeton, NJ 08544, USA}


\begin{abstract}
In this paper we present a catalog of \ebno eclipsing binaries observed during the first two years (26 sectors) of the \tess survey. We discuss selection criteria for eclipsing binary candidates, detection of hither-to unknown eclipsing systems, determination of the ephemerides, the validation and triage process, and the derivation of heuristic estimates for the ephemerides. Instead of keeping to the widely used discrete classes, we propose a binary star morphology classification based on a dimensionality reduction algorithm. Finally, we present statistical properties of the sample, we qualitatively estimate completeness, and discuss the results. The work presented here is organized and performed within the \tess Eclipsing Binary Working Group, an open group of professional and citizen scientists; we conclude by describing ongoing work and future goals for the group. The catalog is available from \url{http://tessEBs.villanova.edu} and from MAST.
\end{abstract}

\section{Introduction}

The Transiting Exoplanet Survey Satellite (\tessnts; \citealt{ricker2015}) was launched in April 2018; during its 2-year prime mission, it monitored $\sim$200,000 bright stars for exoplanets across the sky with a 2-min short cadence. In addition, \textsl{TESS} acquired full-frame images (FFIs) every 30 minutes. \textsl{TESS} is currently in its extended mission, where targets are observed with a 2-min and a 20-sec cadence, and FFIs are acquired every 10 minutes. The primary \tess mission has been to discover and characterize exoplanets, but \tess data enable a much broader swath of science -- all fields that benefit from precise timeseries of bright stars, in fact -- including eclipsing binary systems.

Eclipsing binaries (EBs) serve as one of the pillars of stellar astrophysics. The well-understood laws of motion that govern binarity and the alignment with the line of sight make their analysis a tractable geometrical problem \citep{prsa2018}, yielding accurate masses, radii, temperatures, and luminosities of EB components \citep{torres2010}. Because of that, EBs are used to calibrate stellar models across the Hertzsprung-Russell diagram \citep{Serenelli2021}, rendering them important to essentially every field in astronomy.

Binary stars are ubiquitous: more than half the stars with masses of $1 M_\odot$ or higher are found in binary or multiple systems \citep{raghavan2010,Sana2012,Moe2017}. Binaries are thus a natural product of star formation and make up a large fraction of the visible universe. Understanding binaries means understanding stellar formation and evolution \citep{stacy2010}, internal stellar structure by way of tidal interactions and/or tidally induced pulsations \citep{thompson2012}, accretion physics in semi-detached binaries \citep{bisikalo2010}, and much more. At the same time, there are many open questions that remain, for example what mechanism (or combination of mechanisms) drives multiplicity rates \citep{duchene2013}, what determines the distribution of mass ratios \citep{wells2020}, how does (close) binarity affect stellar evolution of stars (blue stragglers, yellow giants, magnetic interaction; \citealt{mathieu2009}), how does orbital tightening work given that the Kozai-Lidov cycles might not fully explain it \citep{hwang2020}, how do exoplanets that orbit binary stars form and evolve \citep{paardekooper2012}, etc. It is \emph{eclipsing} binaries that hold the answer to these and similar questions. Astronomers have studied EBs for over two centuries, ever since John Goodricke suggested in 1782 that eclipses are responsible for Algol's brightness variation. So why are the answers so elusive?

The principal reason for the remaining open questions is that traditional observational techniques (ground-based photometry, follow-up spectroscopy, long baseline interferometry) are best suited to single objects and are time-consuming. Thus, it is difficult to draw inferences on the entire population of binaries. The first mission that made a significant breakthrough in the EB science was \kepler\ \citep{kirk2016}, but the downside of \kepler\ is that its targets are faint (i.e.~difficult to follow up), limited to a single $\sim$100-deg${}^2$ field, and telemetry and on-board storage did not allow FFIs to be sent back to Earth. Those obstacles are largely overcome by \tessnts: \tess\ observes on the bright end, targets are sourced across the sky, and we have 10-min FFIs. These benefits come at the expense of large (21 arcsec) pixels, adversely affecting crowded fields, but \tess\ still serves as a proverbial gold mine for EBs away from the Galactic plane.

EBs play a less celebrated role in exoplanetary science, where $\sim$40\% of false positives at low Galactic latitudes are attributed to their diluted lightcurves \citep{morton2011}. About 25\% of \tess Objects of Interest (TOIs) that were examined by ground-based photometric follow-up turned out to be background binaries. Thus, having a good census of EBs feeds back to identifying false positives before pointing costly follow-up telescopes in their direction.

In this paper, the first in the \tess EB series, we present a sample of \ebno EBs observed by \tess in the first 26 sectors of observation. In Section \ref{sec:detection} we describe lightcurve detection and identification; in Section \ref{sec:ephems} we explain how ephemerides for each system are determined and refined; in Section \ref{sec:validation} we present an automated data validation pipeline called ICED-LATTE; in Section \ref{sec:props} we focus on statistical properties of the EB sample and qualitatively assess completeness; in Section \ref{sec:catalog} we describe the contents of the catalog; finally, in Section \ref{sec:discussion} we discuss some of the most interesting results and provide a list of ongoing projects and future goals for the working group.

\section{Detection of \tess EBs} \label{sec:detection}

Extracting a sample of EBs from the observations of $\sim$200,000 2-min cadence light curves involved multiple complementing efforts. We describe these efforts here.

\begin{description}
\item[Proposed targets] As part of the \tess Guest Investigator program, we proposed for 3889 targets in Cycle 1 and 3067 targets in Cycle 2. The targets were selected from all public binary star catalogs served on VizieR \citep{ochsenbein2000}, the General Catalog of Variable Stars (GCVS; \citealt{samus2017}), and the Spectroscopic Binary Catalog (SB9; \citealt{sb2004}). The proposed targets were prioritized by a multitude of factors: (1) spatial position in the sky, i.e., the number of visits, (2) $T$ magnitude, as it appears in the \tess Input Catalog (TIC; \citealt{stassun2018}); (3) membership in the Detached Eclipsing Binary Catalog (DEBCAT; \citealt{southworth2015}); (4) classification certainty; and (5) scientific importance. Each criterion was assigned a numerical value (positive or negative) and their sum determined a priority value in the target list. Thus, bright DEBCAT members in the continuous viewing zone were the highest priority targets. Targets on the faint end observed in a single sector and without certain classification were the lowest priority targets. Of the proposed targets, 745 were selected for observations in Cycle 1 and 999 were selected for observations in Cycle 2.


A total of 6699 EB candidates were identified via the Planet Hunters TESS (PHT) citizen science project \citep{eisner2021pht}, which is hosted by the Zooniverse platform \citep{lintott08, lintott11}. The project engages over 30,000 registered citizen scientists in the search for transiting exoplanets in the 2-minute cadence light curve obtained by \textit{TESS}. The identification of eclipsing binaries and multi-stellar systems is a natural by-product of this large scale visual vetting effort. In brief, each 2-minute cadence \textit{TESS} light curve is visually inspected by 15 citizen scientists, who identify the times of any transit-like signals before moving on to the next light curve. Once all of the data from a given \textit{TESS} sector have been classified, the classifications from the individual volunteers are combined using an unsupervised machine learning algorithm. This allows us to identify times of potential transit-like events in each light curve and rank all of the candidates from most to least likely to contain a transit signal \citep[for details see][]{eisner2021pht}. The 500-700 highest ranked candidates per sector are visually inspected by the PHT science team and grouped into `planet candidates', `EB candidates' and `other'. In total, we identified 2720 EB candidates using this methodology. 

In addition to this classification pipeline, each target has an independent discussion forum, where the citizen scientists can discuss the data, and flag the signals to the science team using searchable hashtags. By the end of the primary mission, the tags `EB', `eclipsing binary', or similar versions thereof had appeared over 46200 times on the discussion forums, corresponding to 5759 individual TESS targets. All of these targets were considered as potential EB candidates and were kept for further vetting. A total of 1780 candidates were discovered via both of these methods, bringing the total number of EB candidates identified via PHT to 6699. 

A second manual search for \tess EBs was carried out by another team of seven citizen scientists, the Visual Survey Group (VSG), independently of the PHT effort.  Up until Sector 21, the VSG collectively scrutinized the Candidate Target List light curves \citep[CTL; ][]{stassun2018} which were binned at six points per hour. Data from subsequent sectors were binned at two points per hour. All data were prepared and surveyed with the LcTools software \citep{2021Schmitt} from FITS files stored at the Mikulski Archive for Space Telescopes (MAST).

\item[The Weird Detector pipeline]

\cite{wheeler2019} introduced The Weird Detector, a phase dispersion-based periodic signal detection algorithm with minimal requirements for signal morphology. The merit function ($\zeta$) of a given trial period ($P_{\mathrm{trial}}$) is calculated using (I) local decrease in $\chi^2$ of the binned, phase-folded light curve and (II) the kurtosis characterizing the tailedness of the binned flux distribution to ensure a mostly flat baseline with one (or a few) excursion(s) representing a dimming event. \cite{chakraborty2020} applied the algorithm to 248,000 2-minute light curves from the first 13 sectors of \tess data; given the highly general nature of the algorithm's candidate signal-finding goal---no particular target shape is optimized for in the pipeline's signal-finding process---the relatively higher SNR of eclipsing binaries compared with many other periodic sources makes them a large fraction (313/377) of the candidate signals. All candidates from this pool were manually vetted, ultimately yielding no novel detections, but with an overlap of 265 true-positive signals.

\item[SPOC pipeline]

All TESS 2-minute pixel stamps are processed by the Science Processing Operations Center (SPOC) pipeline \citep{jenkins2016}. The pipeline runs optimal photometric extraction, followed by two types of light curves: Simple Aperture Photometry (SAP) and Pre-search Data Conditioning (PDC) light curves. The SAP light curves are background-corrected, but have no additional detrending, while constructing the PDC light curves includes detrending for common-mode instrumental systematics using co-trending basis vectors empirically calculated from other sources on the corresponding detector \citep{smith2012, stumpe2014}. The PDC light curves are also corrected for flux contamination from nearby stars. Here we chose to use the SAP light curves to avoid cases where the detrending affects the astrophysical signal \citep{stumpe2012,twicken2010,morris2020}.

The SPOC pipeline also searches the PDC light curves for transiting planet signatures using an adaptive, wavelet-based matched filter \citep{jenkins2002,jenkins2010}, and the resulting TCEs are fitted with an initial limb-darkened transit model \citep{li2019} and subjected to a suite of diagnostic tests to help determine whether the transit signature is due to a planet, an eclipsing binary, stellar variability or instrumental effect.

\end{description}

\begin{figure}
    \centering
    \includegraphics[width=\textwidth]{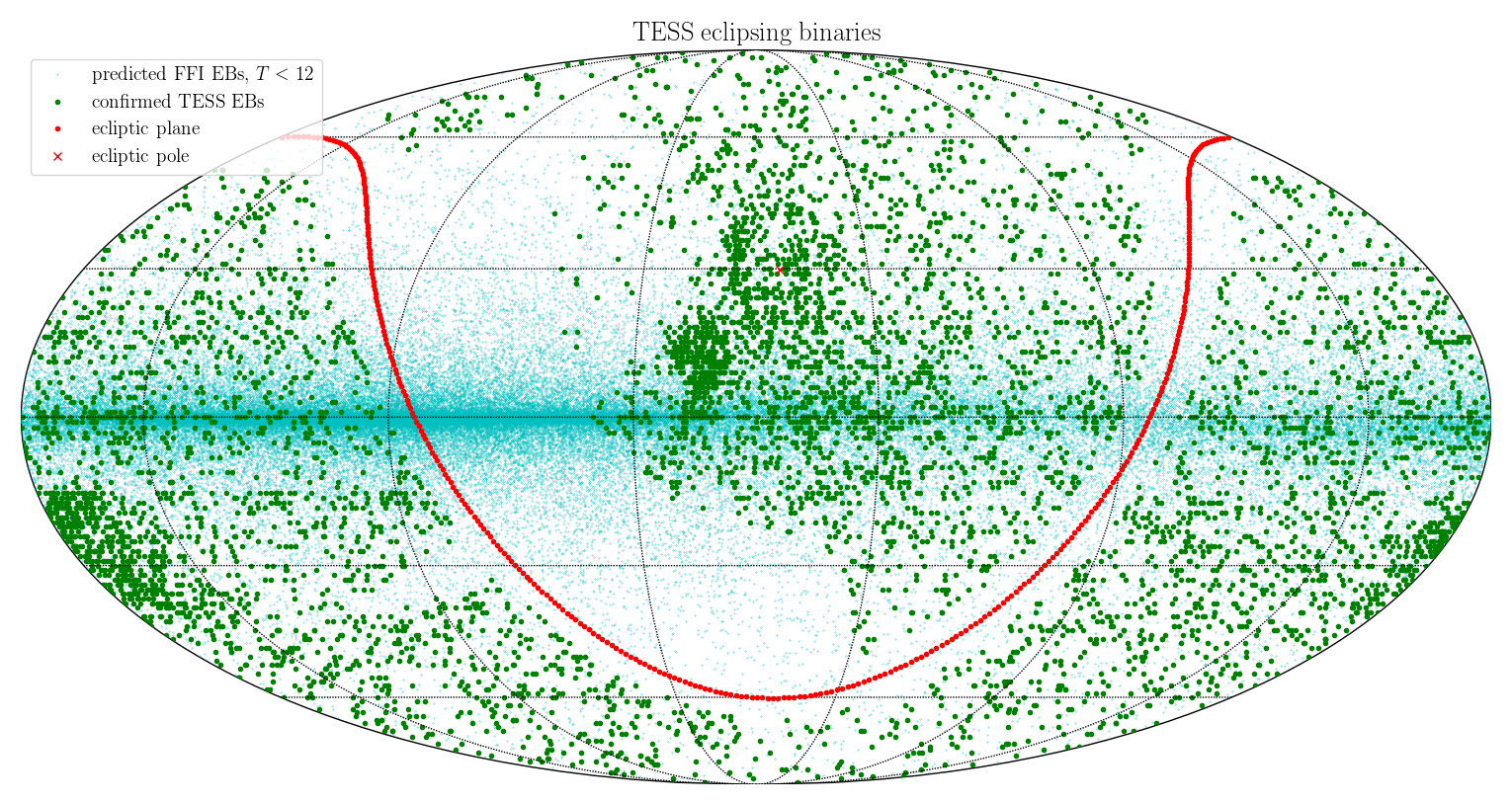} \\
    \caption{Map of \tess EBs observed in sectors 1--26 in the Galactic reference frame. Depicted in green are all vetted and validated EBs observed with the 2-min cadence. Depicted in cyan are the simulated EBs brighter than $T=12$ \citep{wells2021}. The dearth of systems in the region north of the ecliptic plane is due to the change in boresight in sectors 14-16 and 24-26, where the satellite was pointed at $+85^\mathrm{o}$ instead of the nominal $+54^\mathrm{o}$ to mitigate excessive contamination by stray Earthlight and Moonlight in cameras 1 and 2.}
    \label{fig:tess_ebs}
\end{figure}

\begin{figure}
    \centering
    \includegraphics[width=0.9\textwidth]{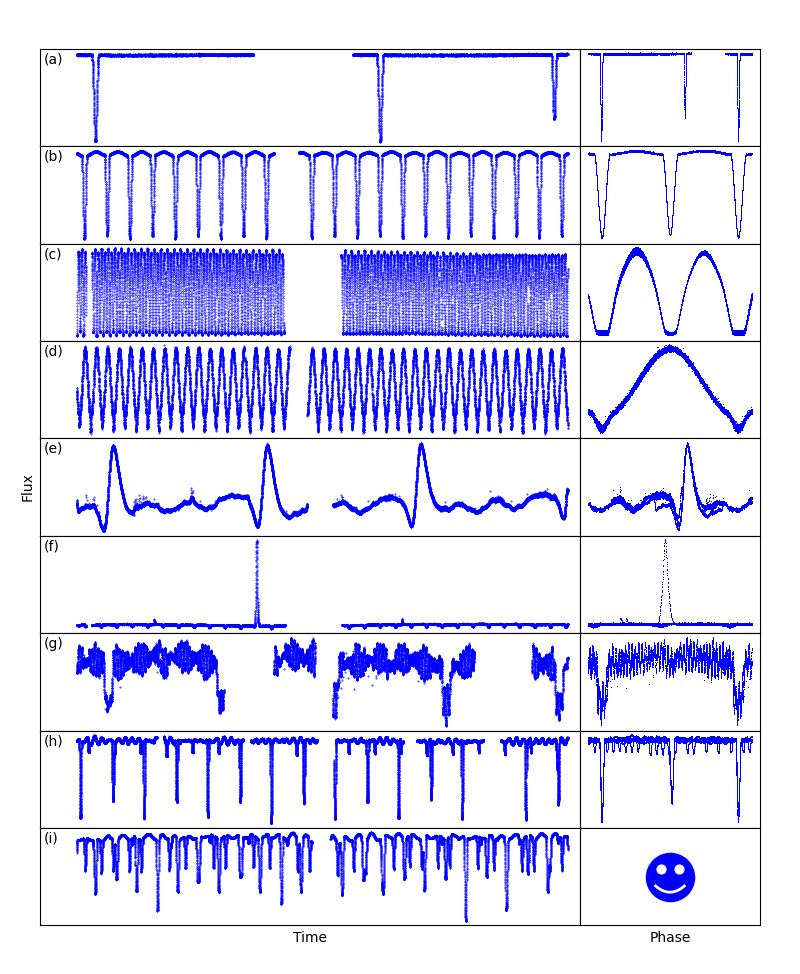} \\
    \caption{A showcase of interesting objects observed by TESS. The panels on the left display timeseries and the panels on the right depict phase plots. The objects are, top-to-bottom: (a) wide eclipsing binary TIC 24935204; (b) benchmark-grade eclipsing binary TIC 33419790; (c) totally eclipsing contact binary TIC 5674169; (d) high reflection subdwarf binary TIC 31690845; (e) eccentric ellipsoidal binary TIC 11046410; (f) flaring M-dwarf pair TIC 436869712; (g) eclipsing binary with a pulsating component TIC 10891640; (h) multiple signal eclipsing binary TIC 375422201; and (i) quadruple star system TIC 424508303.
    }
    \label{fig:showcase}
\end{figure}

\section{The determination of EB ephemerides} \label{sec:ephems}

\subsection{The QATS code}
The Quasi-periodic Automated Transit Search (\texttt{QATS}) is a pipeline originally developed to find planets with transit timing variations in the Kepler data \citep{Carter2013}. It was subsequently updated and revised to be a more general planet and eclipsing binary search tool, as first applied to K2 by \cite{Kruse2019}. The implementation used here is identical to that of \cite{Kruse2019}, so refer there for full details. In brief, \texttt{QATS} models a transit at every cadence in a light curve and compares that model's $\chi^2$ fit to a pure polynomial continuum. It then runs a period folding search over that sequence of $\Delta\chi^2$ to identify periodic signals where the transit fit is better than the continuum. Once it has identified a candidate periodic signal, it runs a more thorough transit fit to accurately measure the transit or eclipse parameters.

\subsection{The ECLIPSR code}

\texttt{ECLIPSR} (\textit{Eclipse Candidates in Light curves and Inference of Period at a Speedy Rate}, \citealt{ijspeert2021}) is an algorithm that operates in two main stages: finding eclipses in the light curve and subsequently determining the periodicity in those eclipses. Finding (individual) eclipses in the light curve is achieved by looking for peaks in its time derivatives. This enables the successful identification of eclipses in light curves that show strong additional (intrinsic) variability compared to the eclipse signal. This process is fully automated and produces a score at the end for each light curve that can be used to separate light curves that show eclipses from those that do not contain an eclipse signal. Here, we start off from a list of pre-determined EB candidates and use the \texttt{ECLIPSR} algorithm mainly for its ability to determine ephemerides. 

\subsection{The BLS run}

The final algorithm employed a traditional and well-tested approach to searching for transit-like signals: the Box Least Squares Periodogram (BLS; \citealt{Kovacs2002boxfitting}). In the BLS periodogram, a sliding box-like signal is passed over the light curve for a range of orbital periods, and the likelihood of the model is recorded at each orbital period. The resulting periodogram is expected to peak at integer multiples of the any box-like periodic signals in the light curve, and thus is well-suited to the detection of eclipses. We used the BLS implementation in Astropy \citep{astropy:2013} with the \textsc{lightkurve} package \citep{GeertBarentsen2019KeplerGO} to normalize and prepare the light curves. No further pre-processing was applied to the light curve beyond a simple normalisation of flux. Although straightforward to implement and run, the BLS algorithm relies on strictly periodic signals and thus can not identify single eclipse events. The orbital periods searched by BLS ranged from 0.1 days to half the time-span of the light curve, with an oversampling factor of 20. 

\subsection{Triage}

All candidate targets were vetted manually by at least one of the authors through a custom web application.  For each TIC entry, phased plots were shown at the period, as well as half-period and double period, for each of the automated ephemeris algorithms discussed above.  

The triage user would then choose which period (if any) correctly represented the signal and classify whether the signal was that of an eclipsing binary or some other variable source.  Subcategories were available to flag eclipsing binaries where the period was ambiguous (whether there are two nearly identical primary and secondary eclipses or no visible secondary eclipse) and whether there was an insufficient number of eclipses in the data to accurately determine the orbital period.

Ephemerides for a single TIC target with periods (or period multiples) within 1\% were automatically flagged as representing the same underlying signal.  All other multiple ephemerides classified as eclipsing binary signals were then treated independently as either blended EBs or true hierarchical systems.

A total of 27,496 candidate ephemerides (from 10,477 unique TICs) and their half- and double-period counterparts were triaged.  4592 ephemerides (from \ebno unique TICs) were manually classified as likely being caused by an eclipsing binary signal and were passed on to the ephemeris refinement algorithm described below.  Additionally, 520 input ephemerides (from 457 unique TICs) were classified as likely eclipsing binaries but with an insufficient number of eclipses to determine an orbital period.  1872 input ephemerides (from 1725 unique TICs) were flagged as requiring further follow-up to determine whether they were eclipsing binaries or pulsating stars (due to sinusoidal signals), 9434 input ephemerides (from 6029 unique TICs) were classified as having no eclipsing binary signal, and 11,078 input ephemerides were marked as duplicates of another ephemeris entry for the same TIC.

\subsection{Refinement and heuristic error estimates} \label{sec:refinement}

To further refine the ephemerides determined from triage, we fit analytical models to the phase-folded light curves and sample the posteriors of the period used for phase-folding and the model parameters. The two analytical models used are \twog \citep{mowlavi2017} and \polyfit \citep{prsa2008}. 

\begin{figure}
    \centering
    \includegraphics[width=0.9\textwidth]{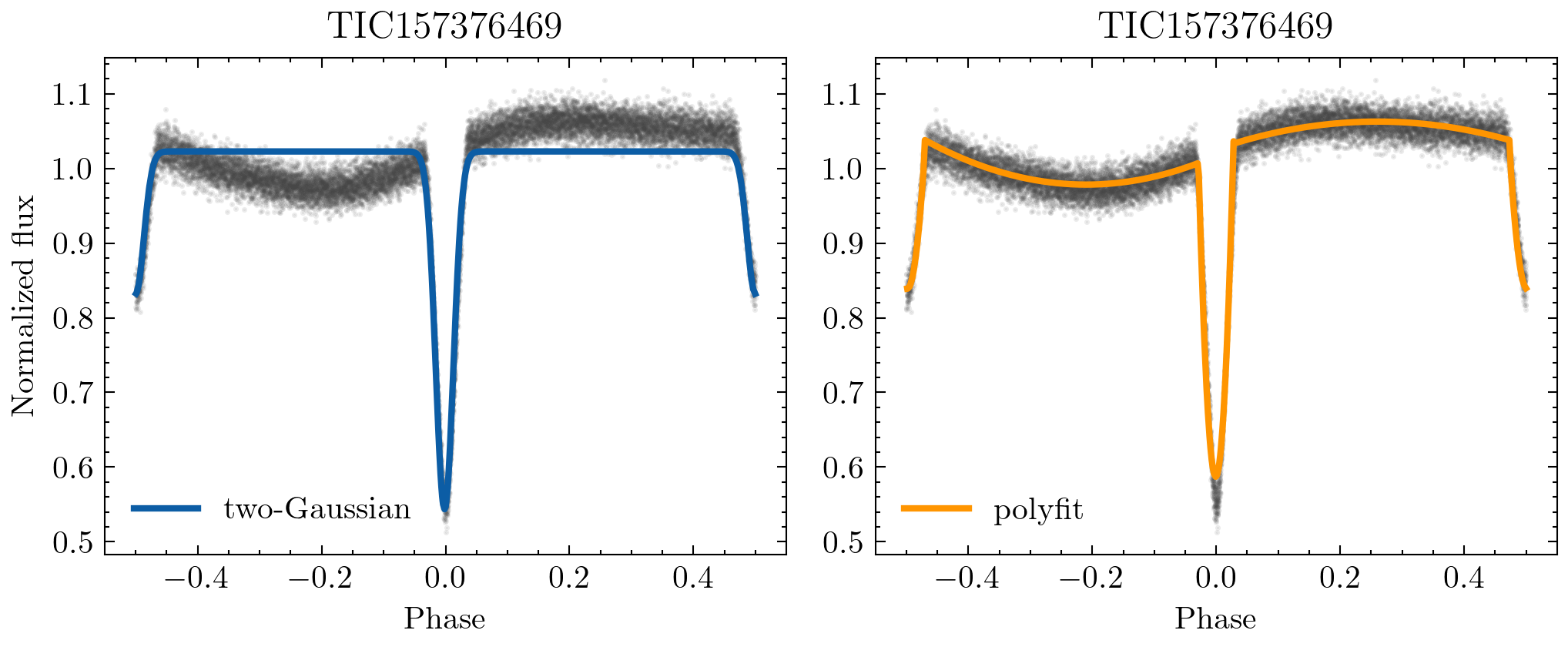}
    \includegraphics[width=0.9\textwidth]{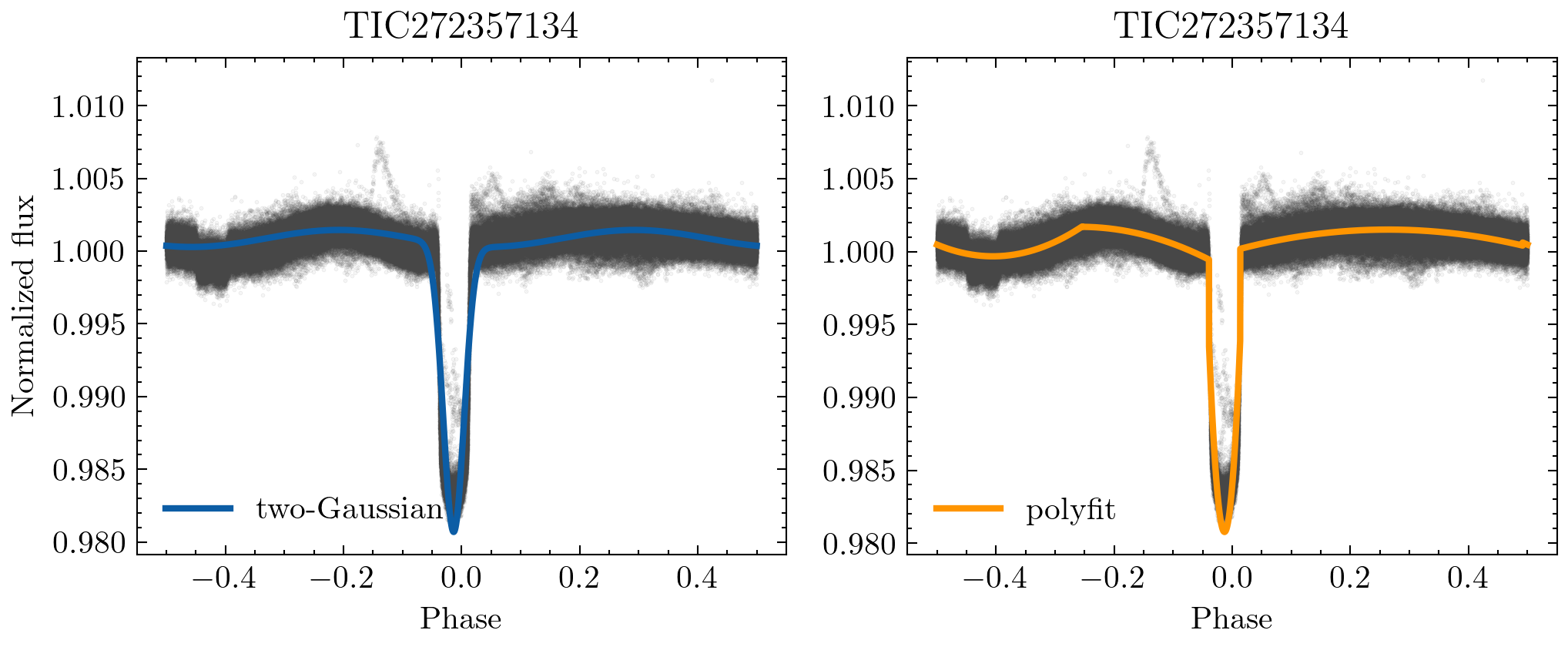}
    \caption{Demonstration of a successful (top panels) and unsuccessful (bottom panels) fit with the two models. The top panels show a system with asymmetric variability outside of eclipse, which cannot be modeled by \emph{two-Gaussians}. However, this does not prevent the model from fitting the eclipses correctly. \emph{Polyfit}, due to its higher degree of freedom, manages to capture both the eclipses and variability outside of eclipse. The bottom panels show the deficiencies of the two models that cause it to not correctly fit the shallow secondary eclipse: the \twog model considers it part of the asymmetric variability outside of eclipse that it is unable to model, while \polyfit fits the variability as an eclipse instead.
    \label{fig:twogpf}
    }
\end{figure}

The \twog models fit a phased light curve by using one or two Gaussian functions and/or a cosine function with its maximum coinciding with one of the eclipses. The Gaussians are supposed to fit eclipses and the cosine function is supposed to fit symmetric ellipsoidal variability. The light curve is fit by all combinations of functions and it finds the solution that minimizes the degrees of freedom while retaining a satisfactory fit. The model with the highest Bayesian information criterion (BIC) is chosen as the best fit and used in the subsequent analysis. The left panels in Figure~\ref{fig:twogpf} depict an example of a \twog fit.

The \polyfit analytical model relies on fitting a piecewise-connected chain of polynomials to the phased light curve. The constraints imposed on the chain are that it should be connected and smoothly wrapped in phase space. There is no requirement that the chain be differentiable in the knots, which allows it to easily fit the discrete breaks caused by eclipses. As such, the knots are typically positioned at the ingress and egress of the eclipses. We use four quadratic polynomials connected at four knots.  The right panels in Figure~\ref{fig:twogpf} depict an example of a \polyfitnts.

The light curve geometries induced by the eclipses, ellipsoidal variability, spots, and other potentially present signals, are in reality more complex than those that can be modeled with these analytical functions. However, they are sufficient for this preliminary analysis which focuses on period refinement and simple geometrical parameter estimates.

To estimate the initial distributions for the model parameters, we first fit the model to the phase-folded data. For the \twog model, the best fitting combination of eclipses and a cosine term is chosen and its corresponding parameters passed on to a Markov Chain Monte Carlo (MCMC) search with walkers initialized in a tight ball around the fitted parameter values. If there are no eclipses or ellipsoidal variability detected by \twogs (which results in a constant function fit), the light curve is not passed on to MCMC and flagged as ``failed \twognts". Similarly, we fit an initial polyfit to the phase-folded light curve, and if successful, we initialize a sample around the fitted knot positions and polynomial coefficients. A failed \polyfit is rare, but if the algorithm raises any errors that do not result in a fit, we flag that system as a ``failed \polyfitnts".

We run MCMC with \textsc{emcee} \citep{dfm2019} on each candidate EB light curve with 96 walkers for 2000 iterations. The burn-in typically takes under 100 iterations, but we discard the first 1000 iterations for the mean and standard deviation computation of the parameter posteriors. We also check for potential multiple ``branches" of walkers at different log-probabilities and only use the one with lowest log-probability of the posteriors. Out of the 4584 EB candidates, 98.08\%  were successfully fitted with a \twog model and 99.85\% with \polyfitnts. 99.24\% out of the fitted \textit{two-Gaussians} and 98.89\% of the fitted \textit{polyfits} are within 1\% of the triage period. Additionally, 98.93\% of fitted \twogs are within 1\% of the \polyfit periods.

As depicted in Figure~\ref{fig:twogpf}, \twogs and \polyfit have their deficiencies which hinder their application to all systems. Fortunately, they tend to complement each other (unlike \twognts, \polyfit can fit a wide range of out-of-eclipse signals, including highly asymmetric ones; while \twogs will tend to fit eclipses even in noisy data, where \polyfit underperforms). We also run additional checks to ensure that a fitted Gaussian or polyfit truly fits an eclipse and not some other feature of the light curve: 
\begin{itemize}
    \item overlapping eclipses check - ensures that a single eclipse is not fitted by two Gaussians or two polynomials;
    \item noise check - ensures that a Gaussian is not fitted to the data noise;
    \item ELV check - ensures that a Gaussian is not fitted to ellipsoidal variations (ELVs) or another out-of-eclipse variability;
    \item \textit{\polyfit only}: check that the eclipse coincides with a polynomial minimum and not a knot position. If a knot exists with a lower flux value than the polynomial minimum, the eclipse is discarded.
\end{itemize}

Where any one of these checks fails, the eclipse parameters are not reported, even if a \twog or \polyfit model solution exists.

\begin{figure}
    \centering
    \includegraphics[width=0.9\textwidth]{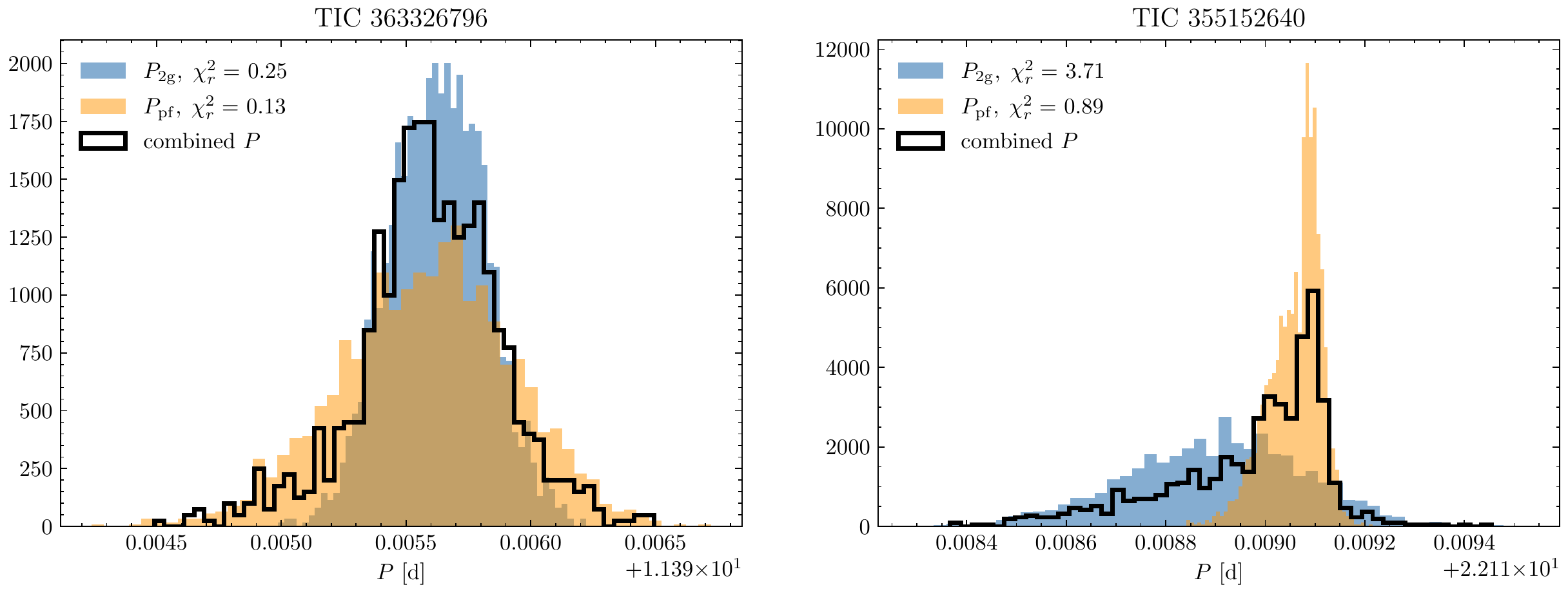}
    \caption{Combined period distribution for a \twog and \polyfit with comparable $\chi^2_r$  (left) and one where \polyfit is significantly better (right).}
    \label{fig:combined_period}
\end{figure}

To fully utilize the information returned from both models and their period and parameter distribution, we compute combined ephemerides, which are ultimately reported in the catalog. The combined ephemerides are computed by sampling both the \twog and \polyfit distributions, under the constraint that the number of samples drawn from each distribution is inversely proportional to the reduced $\chi^2$ value ($chi^2_r$) of the mean fit (Figure~\ref{fig:combined_period}). The mean and standard deviation of the new combined distribution are then chosen as the final ephemerides. We perform an additional check to ensure that the models whose distributions we sample are reasonable. If a model reports uncertainty of the time of superior conjunction larger than half the fitted period, the time of superior conjunction for that model is discarded. This happens more often in the case of \polyfitnts, as the eclipse model has more degrees of freedom.

\subsection{Morphology classification} \label{sec:morph}

For morphological classification of the light curves, we rely on the morphology parameter as defined in \citet{matijevic2012}. The morphology parameter is defined as a continuous variable on the range [0,1], where 0 corresponds to the widest detached systems and 1 to ellipsoidal variables. We use the \kepler EBs dataset to train a neural network (NN) model on the light curve geometry and output a value of the morphology parameter. The \kepler and \tess EBs datasets are both preprocessed in the same way: phase-folded with the final catalog ephemerides, and binned in 1000 phase bins over the range [-0.5, 0.5]. The NN architecture consists of an input layer of size 1000, corresponding to the phased light curves; three dense layers of size 300, 100 and 30, respectively; and an output layer of size 1, corresponding to the morphology parameter. The NN is trained on 60\% of the \kepler data set and tested on the remaining 40\%, yielding a mean squared error loss of 0.0035.

\begin{figure}
    \centering
    \includegraphics[width=0.8\textwidth]{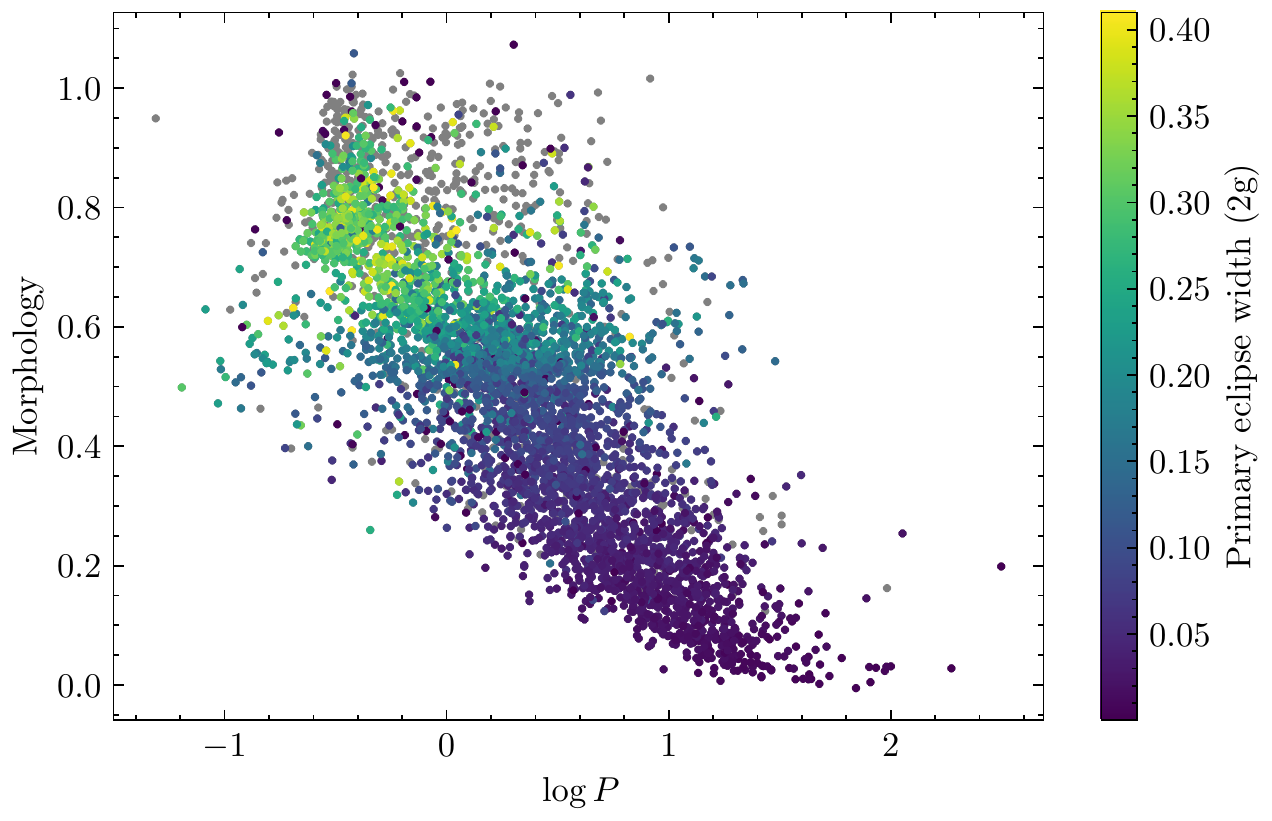}
    \caption{The relationship between morphology parameter, logarithm of the period and primary eclipse widths, as determined from the \twog model.
    \label{fig:logp_morph}
    }
\end{figure}

Figure~\ref{fig:logp_morph} demonstrates the relationship between the morphology, periods and primary eclipse widths in the catalog. Long-period systems have a morphology parameters close to zero, signaling wide detached binaries, while the shortest period systems tend towards morphology of 1, which corresponds to contact binaries and ellipsoidal variables. The primary eclipse widths, as determined by the \twog model, are encoded in the color map, further confirming that the morphology parameter captures the expected variability in eclipse widths: narrow eclipses in systems with to morphology close to 0 and wider eclipses going towards morphology of 1.

\section{EB validation} \label{sec:validation}


Data validation reports, for the purposes of this project, are cumulative accounts of analyses carried out using \textit{TESS} data for each individual target flagged as an EB. Each report contains, but is not limited to, the full lightcurve, target parameters, background flux plots, aperture size plots, power spectra, evolutionary tracks, and results from several eclipsing binary data validation tests. The purpose of these validation reports is to support the eclipsing binary classifications of the \textit{TESS} EBs working group. 

Eclipsing binary data validation resulted primarily from reformatting the interactive code, LATTE \citep[Lightcurve Analysis Tool for Transiting Exoplanets; ][]{eisner2020latte}. LATTE's initial purpose involved identifying false positives in exoplanet classifications due to instrumentation and astrophysical productions of exoplanet-like signatures. The development of ICED LATTE (Interlacing Code for Eclipsing binary Data validation) redirected the purpose of the analysis tool to validate classifications of \textit{TESS} EBs and flag the presence of false positives. Additionally, {\sc LATTE} was modified to work without its interactive plots with the aim of producing quick validation reports for the $\sim$5000 classified EBs. 
Additions to ICED LATTE that were not in LATTE include numerous EB validation tests designed to identify whether the signal is on target, tests to determine whether the signal favours the planet or the EB scenario, and a generated likelihood that the target is an EB.

\begin{figure}
    \centering
    \includegraphics[width=\textwidth]{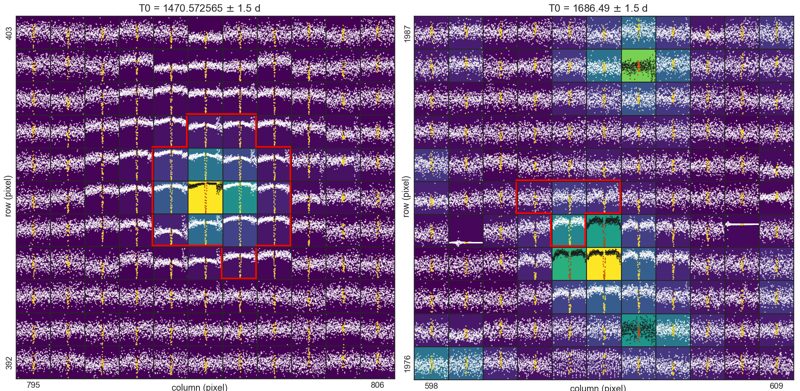} \\
    \caption{A light curve extracted for each pixel around the target star, where the red line indicates the aperture used to extract the 2-minute cadence SPOC light curve. Left: example of an EB which is on target (TIC 255700967). The overall EB score for this target is 0.87. Right: example of a signal that is not on target (TIC 230530979). The overall EB score for this target is 0.42.}
\end{figure}

A significant portion of the validation tests are based on lightcurve analysis. However, the most telling of the validation tests is likely the centroid movement test. The large, 21 arcsec pixels substantially increase the probability of background contamination from sources other than the \textit{TESS} target being observed. In the event that an EB is captured in the same aperture as the target, the target itself may appear to be an EB from lightcurve analysis. The centroid movement test identifies the presence of other objects in the aperture of the target and measures any movement of the central point of light. If this centroid appears to move slightly between the background object and the target, then one of the two is likely an EB. This is due to the change in flux from the EB. During times of eclipses, the EB appears dimmer and the centroid of light moves toward the stable-flux object. Intuitively, during non-eclipse times, the EB appears brighter and the centroid of light moves toward the EB. By analyzing this centroid movement, we are able to determine which of the objects is the true EB and thus confirm or dismiss the EB classification of the \textit{TESS} target.

In addition to results from the validation tests, a rough likelihood that the target is an EB is also calculated and included in the data validation report. It should be noted that these likelihoods are extremely general in nature and only reflect the results of the validation tests which are run on the targets. Each test results in a numerical value between 0 and 1; 0 is generally assigned to the test result ``Unlikely EB", 0.5 is generally assigned to ``Possible EB", and 1 assigned to ``Probable EB". In brief the validation tests that contribute to this score consist of: 

- \textbf{Centroid motion. } The centroid test makes use of the open-source python package {\sc contaminante}, which executes a pixel-level modeling of the TESS Target Pixel Files to determine the most likely location of the source of the eclipses. The score for this test is scaled inversely with the distance between the location calculated source of signal and the location of the target star. 

- \textbf{Contamination. } The amount of contamination to the TESS aperture from nearby stars. This provides an indication of how crowded the field is and the likelihood of the signal originating from a nearby companion star. The score for this test is scaled inversely with the contamination of nearby source.

- \textbf{Out of transit variability. } EBs with orbital periods shorter than around 3 days are expected to show out of transit variability. As such, we search for this variability for the short period candidates. The significance of such a detection is proportional to the score given for this test. Candidates with periods greater than 3 days are given a score of 0.5 for this test by default. 

- \textbf{Archival classification. } Candidates that have previously been listed as an EB on Simbad are awarded a score of 1 while all other candidates are awarded a score of 0.5.

- \textbf{TCE/TOI. } The TESS automated search pipeline flags lightcurves containing a periodic signals, including both planetary and stellar, as Threshold Crossing Events (TCEs) and TCEs that pass a large number of rigorous planet vetting tests are promoted to TOI status. As such, candidates that are TOIs are given a score of 0; candidates that are TCE's but not TOIs are given a score of 0.75, and candidates that are neither TOIs nor TCE's a score of 0.5. 

\begin{figure}
    \centering
    \includegraphics[width=0.8\textwidth]{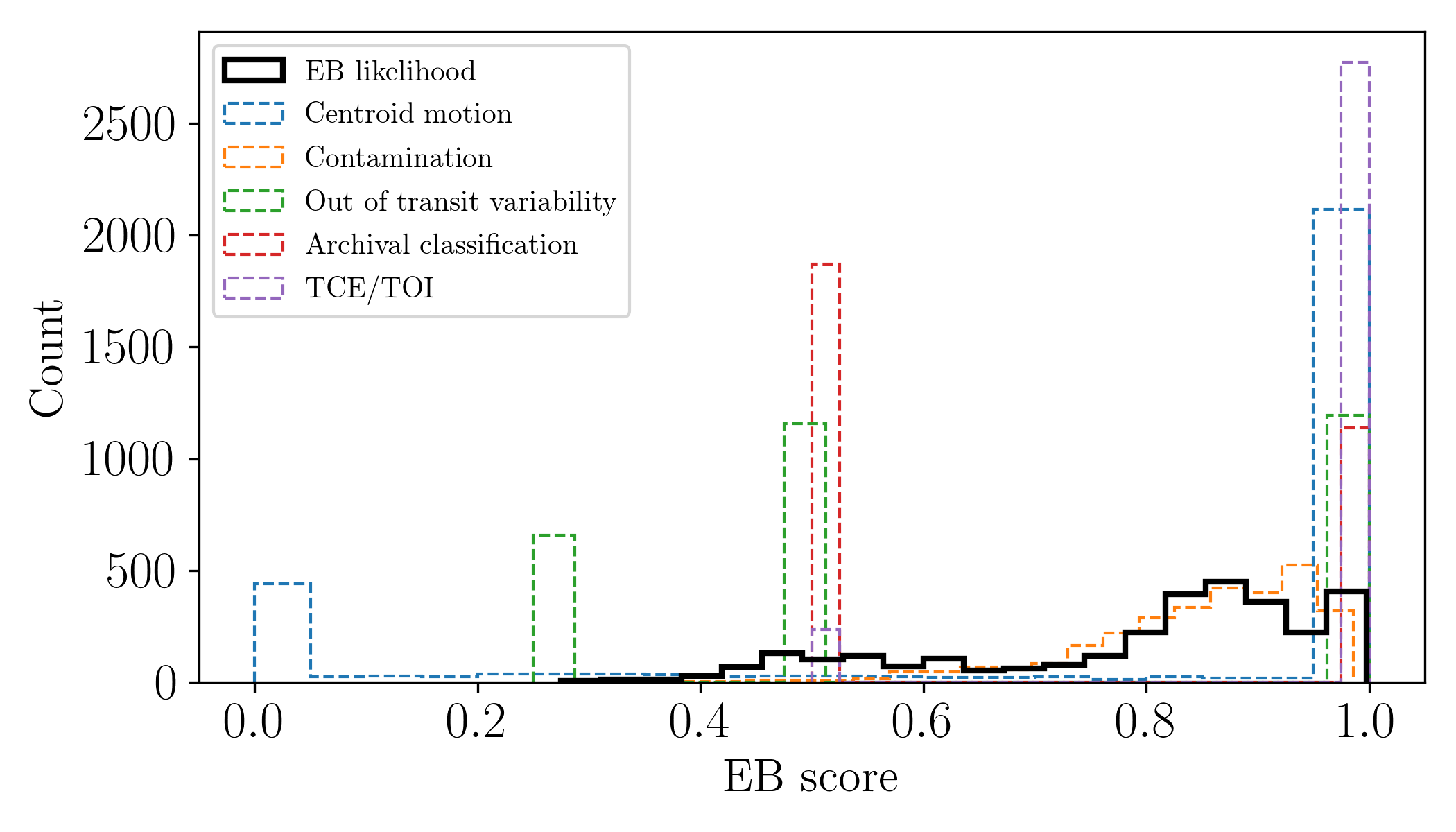} \\
    \caption{Histogram showing the distribution of the individual test scores (dashed outlines) ,which combined give the overall likelihood of the candidate being a real EB (solid black outlines).}
    \label{fig:LATTE_test_result}
\end{figure}

Due to the large number of signals that are off target, we found the centroid motion test to be the most informative to determine whether the EB signal was real. As such, the final EB likelihood is calculated as the average between all of these test scores, with the centroid motion test given a factor of 3 extra weight. The resulting distribution of the final EB likelihood and of all of the individual test results in shown in Figure~\ref{fig:LATTE_test_result}.

\section{Statistical properties of the EB sample} \label{sec:props}

While non-uniform and magnitude-limited, the sample of \ebno EBs observed by \tess allows us to do preliminary bulk analysis on the data-set.

\begin{description}
\item[Orbital periods] the distribution is largely affected by the temporal coverage of observations. Fig.~\ref{fig:logp} shows a bimodal distribution, with a narrow peak at around 0.25 days and a broad peak at around 3 days. The narrow peak corresponds to contact binaries, while the broad peak corresponds to close, detached EBs. We discuss this further in Section \ref{sec:discussion}. The shades of blue correspond to the number of sectors that the targets were observed in; it shows the comparative rates of EBs in the continuous viewing zones (lightest shade) and single sectors visits (darkest shade); the histogram is stacked, i.e.~the values per bin are cumulative over all sector visits.

\begin{figure}
    \centering
    \includegraphics[width=\textwidth]{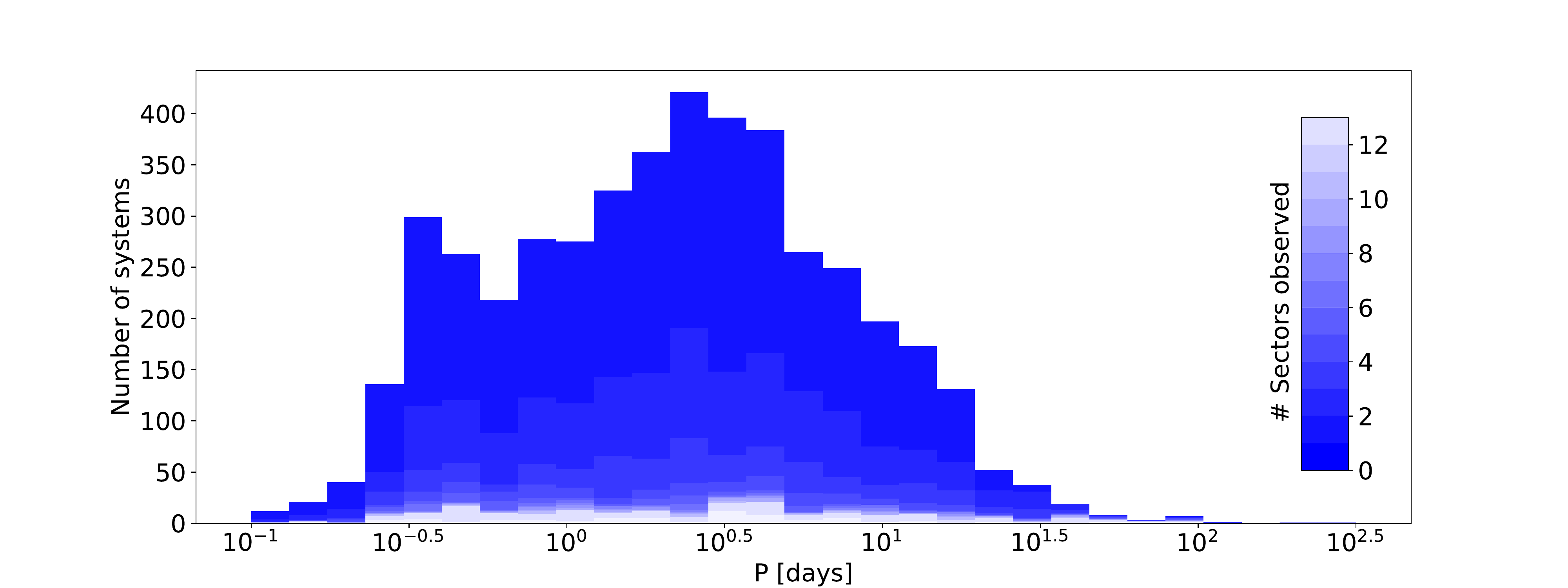} \\
    \caption{The distribution of orbital periods for \ebno EBs observed in sectors 1 through 26. Shades correspond to the number of sectors the target was observed in.}
    \label{fig:logp}
\end{figure}

\item[Eclipse depths] Fig.~\ref{fig:depths} depicts their distribution, along with the distribution of eclipse depth ratios. By definition, the primary eclipse depth is deeper, so we expect the distribution to have a longer tail. The systems with shallow primary eclipses and undetectable secondary eclipses enhance the first bin of the primary depth distribution.

\begin{figure}
    \centering
    \includegraphics[width=\textwidth]{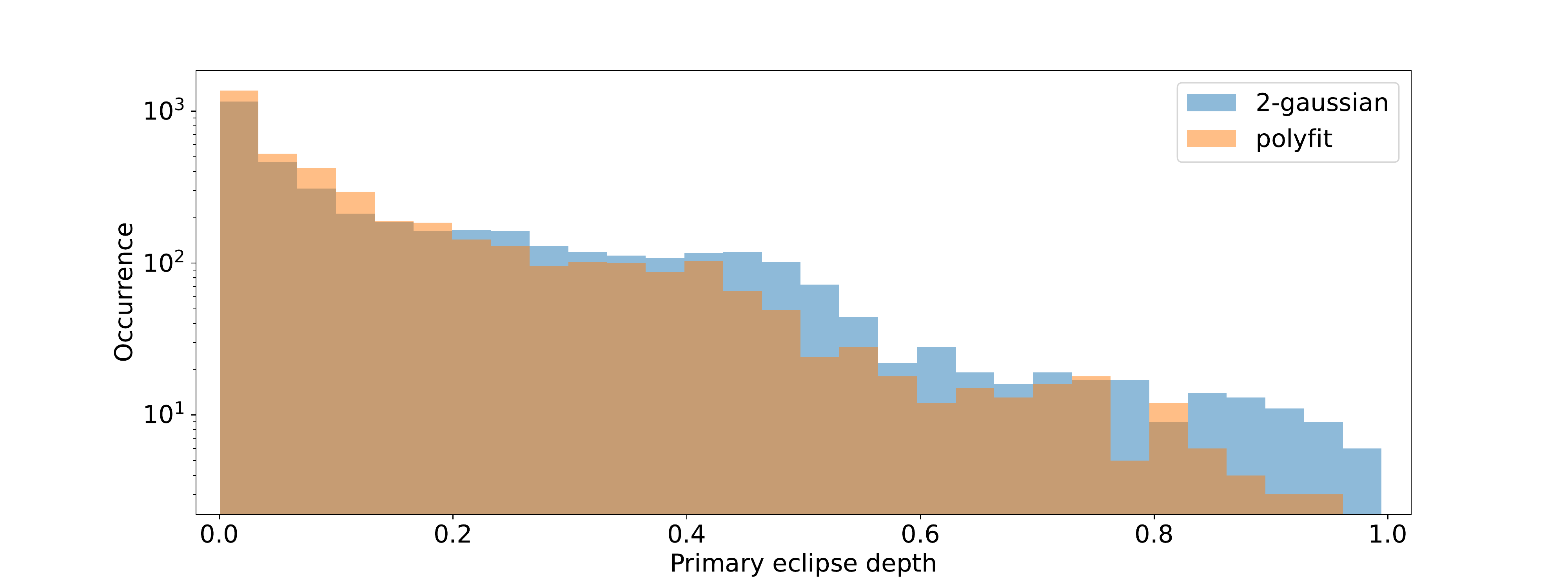} \\
    \includegraphics[width=\textwidth]{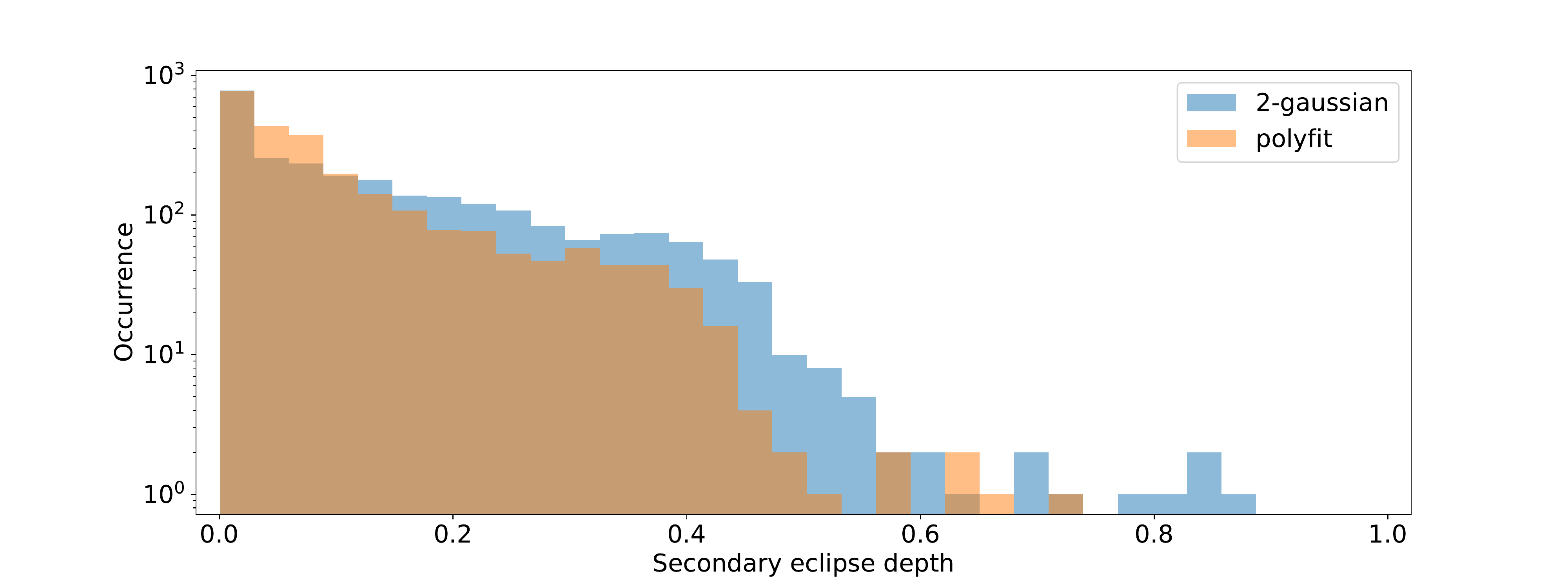} \\
    \includegraphics[width=\textwidth]{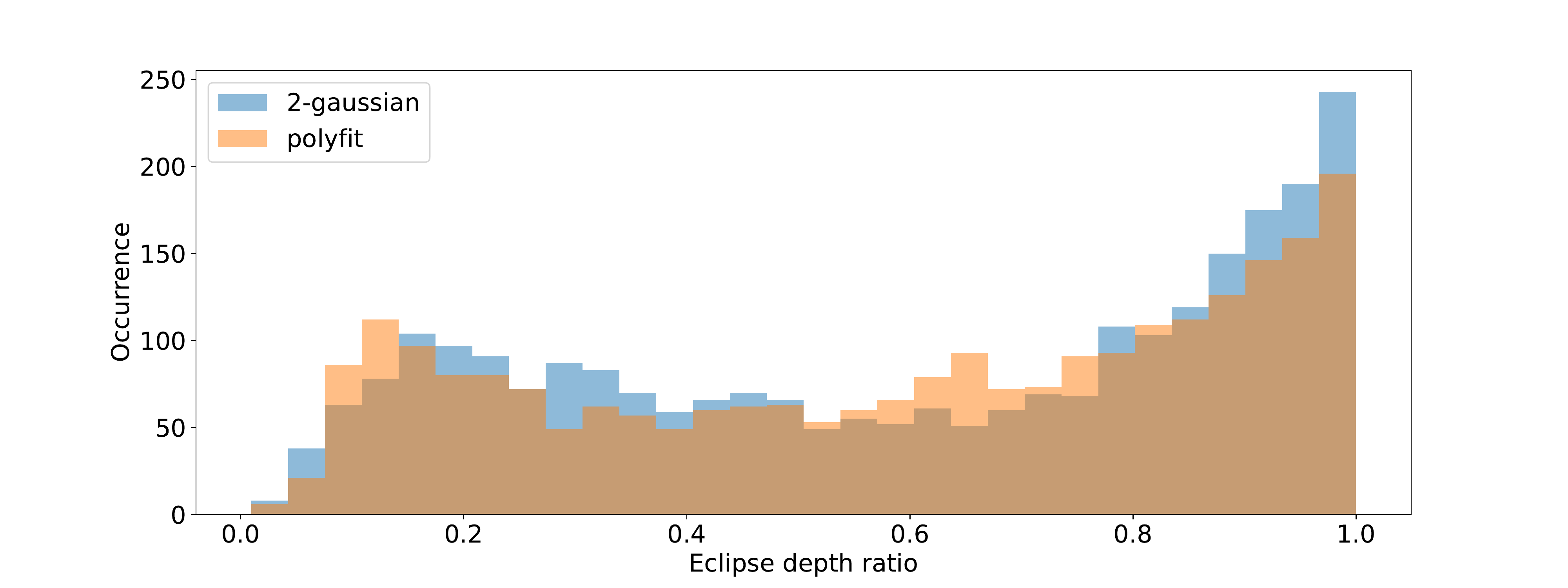} \\
    \caption{The distributions of primary and secondary eclipse depths, and their ratios, derived from the polyfit and 2-Gaussian models. Note the log scale of the $y$-axis. Eclipse depth ratios serve as rough proxies for the surface brightness ratios.
    }
    \label{fig:depths}
\end{figure}

Depth ratios depend on the surface brightness ratio of the two components, and on eccentricity. Surface brightness itself depends on many second-order effects such as gravity darkening, limb darkening, and reflection. Thus, relating eclipse depth ratios to underlying parameters is complicated, but in the broadest sense, the distribution of the depth ratios is aligned with the ratios of effective temperatures. The distribution is bi-modal, with a peak at 1, corresponding \rev{predominantly\footnote{While surface brightness ratio of $\sim$1 can in principle be achieved by luminosity ratios $\neq 1$ and, correspondingly, radius ratios $\neq 1$, the majority of the systems where surface brightness ratio is $\sim$1 corresponds to the same luminosity class, i.e.~twin stars (cf.~Fig.~\ref{fig:gaiacmd2}).} to the equal luminosity class} pairs, and broad, near-flat distribution between 0.2 and 0.8, corresponding to evolved components.

\item[Eclipse separations] phase-space separations of eclipses are a direct measure of the tangential component of orbital eccentricity, $e \cos \omega$. It is reasonably well determined because eclipse positions could be measured reliably. Fig.~\ref{fig:seps} depicts their distribution; note that the $y$ scale is logarithmic. The strong peak at 0.5 corresponds to circular orbits, which are expected to dominate the EB population at these relatively short periods.

\begin{figure}
    \centering
    \includegraphics[width=\textwidth]{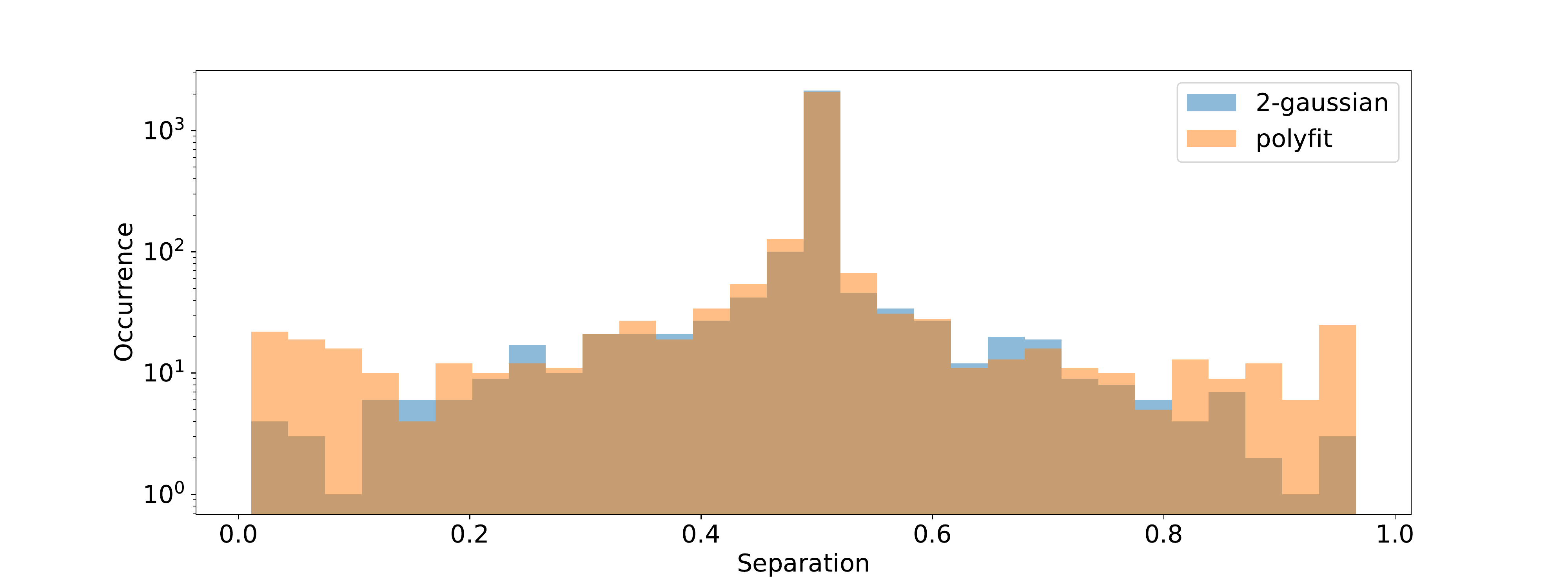} \\
    \caption{The distribution of eclipse separations in phase space. Eclipse separations serve as proxies for the tangential component of eccentricity, $e \cos \omega$.}
    \label{fig:seps}
\end{figure}

\item[Eclipse widths] the least well determined parameters because of their pronounced sensitivity to the exact points of ingress and egress, it is a measure of the radial component of eccentricity, $e \sin \omega$. Fig.~\ref{fig:widths} depicts their distributions. The differences between the two models, polyfit and 2-Gaussian, are most pronounced in eclipse widths; nevertheless, the correlation between the two (bottom panel of Fig.~\ref{fig:widths}) shows the expected trend (distribution around $y=x$) with an expected scatter due to eccentricity effects.

\begin{figure}
    \centering
    \includegraphics[width=\textwidth]{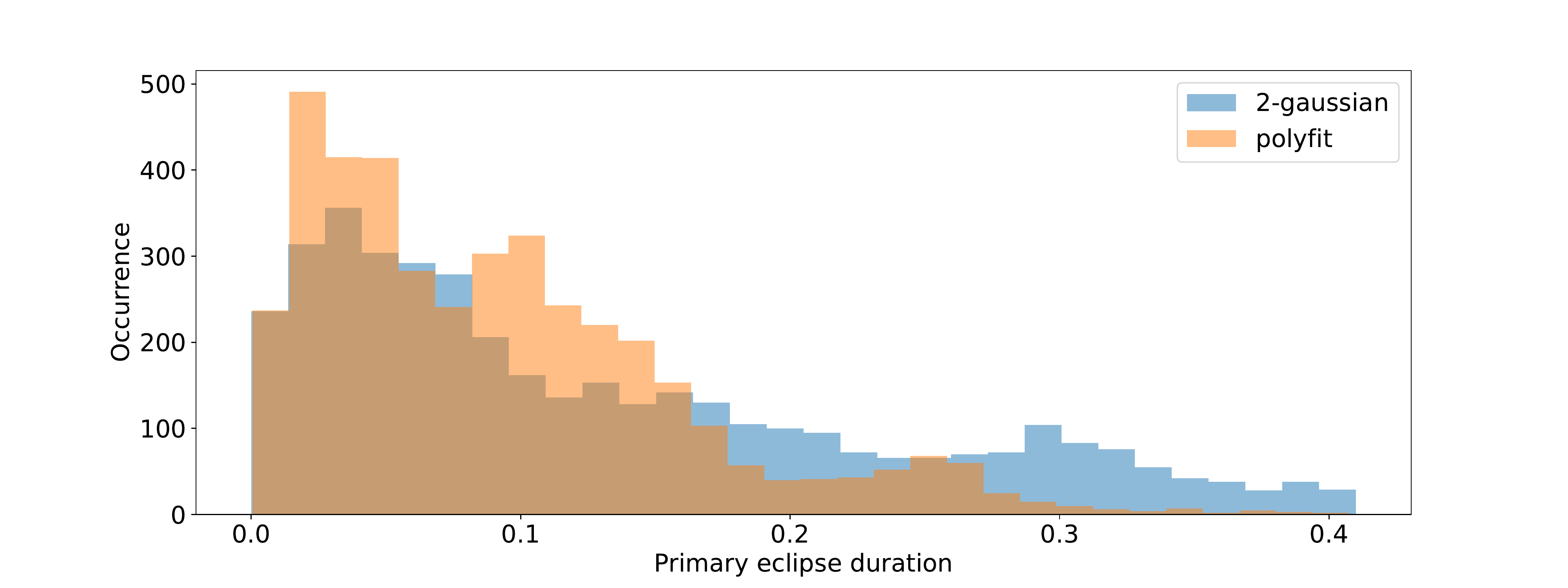} \\
    \includegraphics[width=\textwidth]{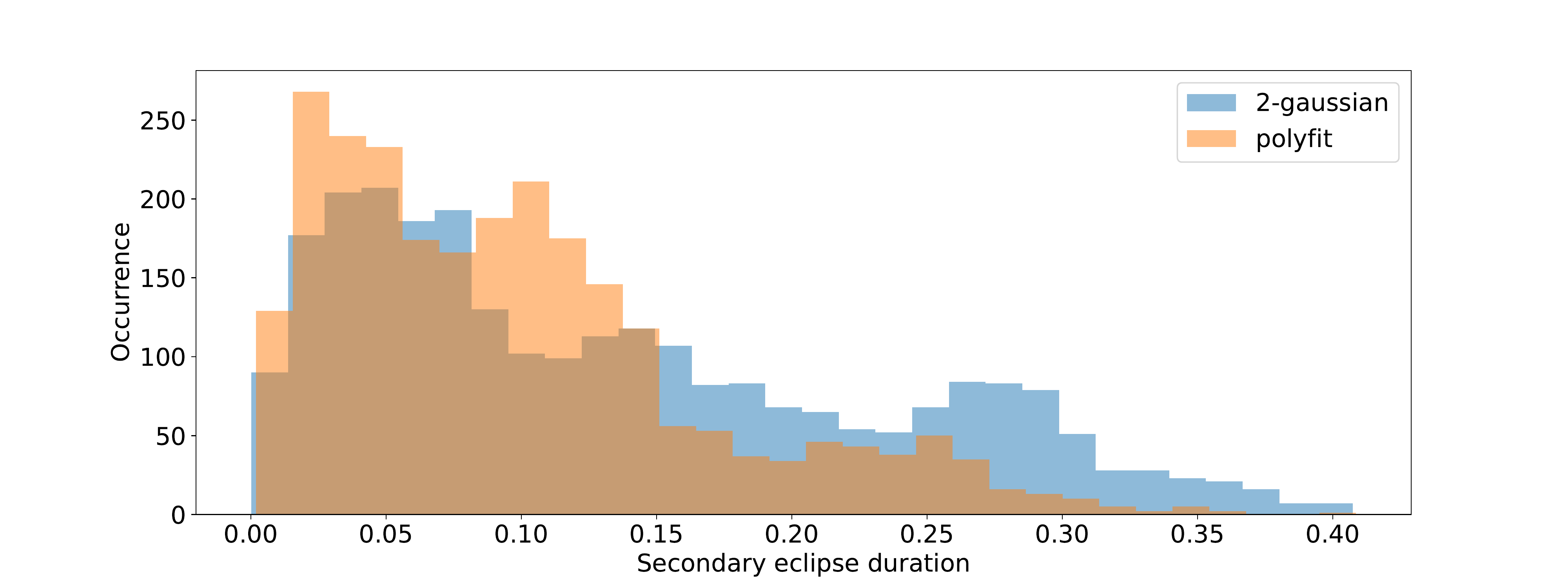} \\
    \includegraphics[width=0.375\textwidth]{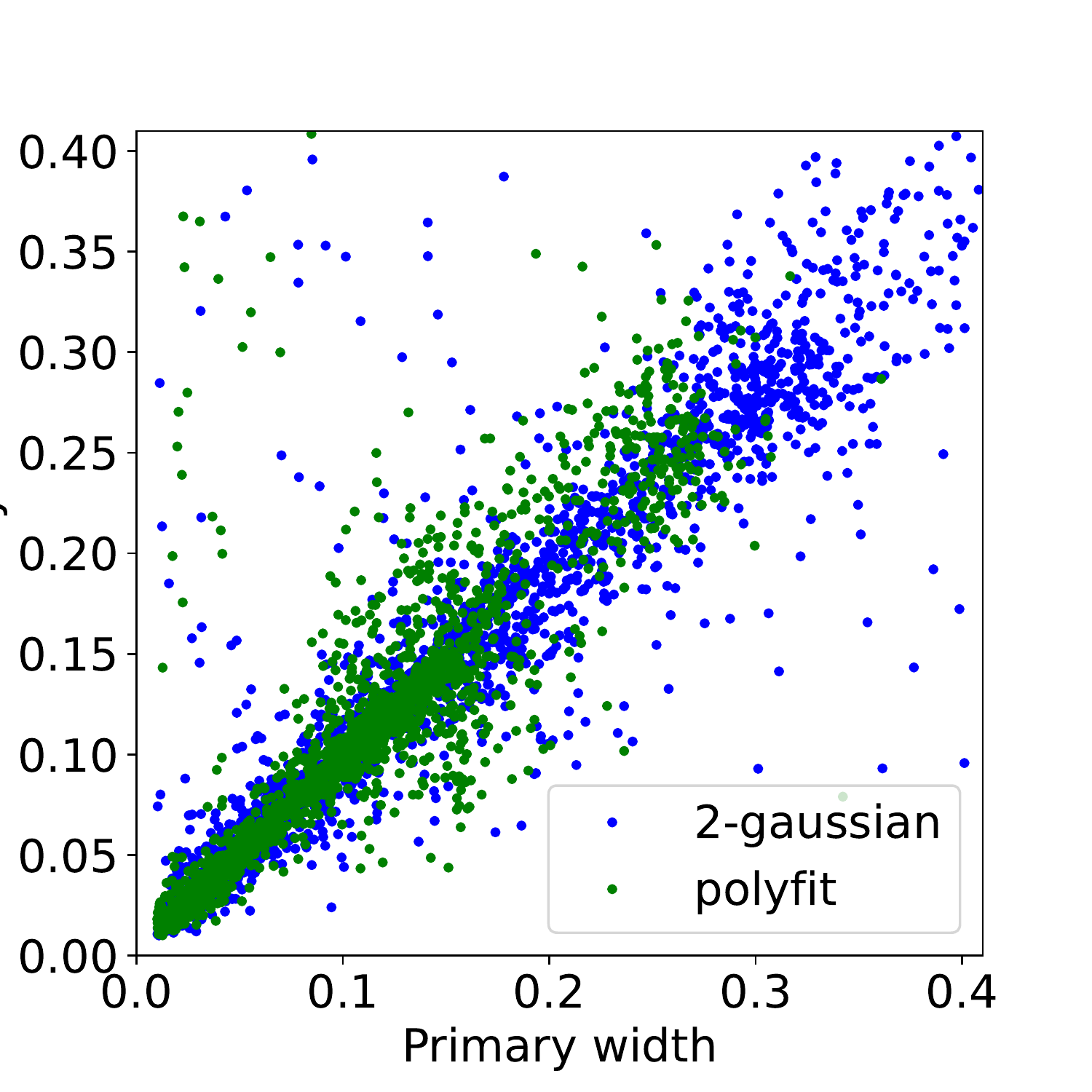} \\
    \caption{The distributions of primary and secondary eclipse widths, and their correlation, derived from the polyfit and 2-Gaussian models. Eclipse widths serve as proxies for the radial component of the eccentricity, $e \sin \omega$.}
    \label{fig:widths}
\end{figure}

\end{description}

\subsection{Catalog completeness}

The sample of EBs presented in this paper is inherently non-uniform: mission targets are selected to maximize the exoplanet detection yield, and guest investigator targets are selected through proposal competition, across a great many science goals. It is thus impossible to use these data to get statistically representative distributions of parameters; we defer that goal until the time when we have a sample of detected and characterized EBs from full-frame images. Thus, when we say ``catalog completeness,'' we mean the overall EB detection success \emph{in the data-set of 2-min cadence target observations}. Quantifying completeness properly thus remains beyond the scope of this paper, but we provide qualitative estimates here.

\begin{figure}
    \centering
    \includegraphics[width=\textwidth]{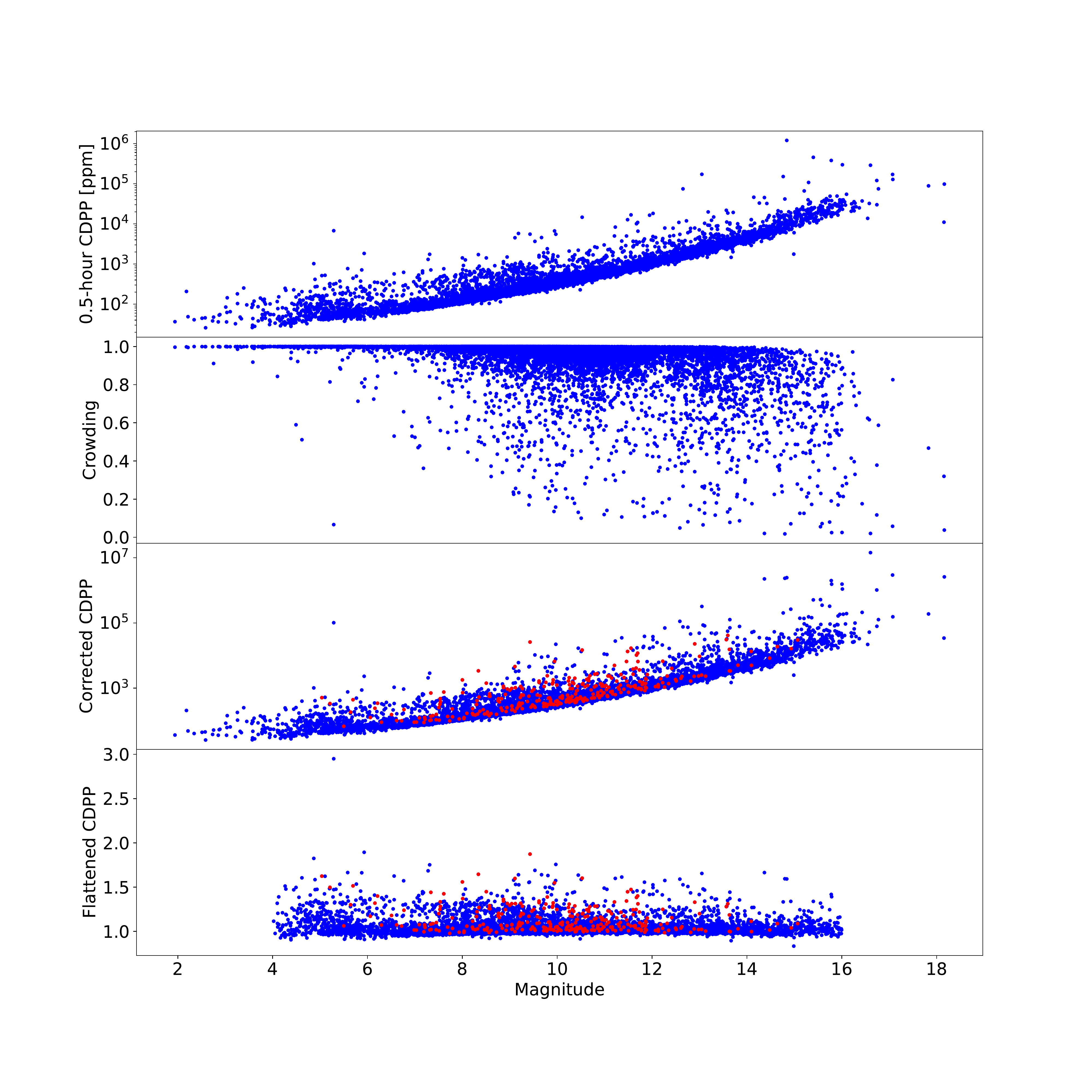} \\
    \caption{The \tess Combined Differential Photometric Precision (CDPP). The top panel depicts the measured CDPP with a 0.5-hr running average, in ppm. The second panel provides a crowding measure per target, i.e., the fraction of light in the aperture from the observed target. The third panel corrects the original CDPP for crowding: corrected CDPP = CDPP / crowding. This assumes that the contribution of background light does not vary with time, which may or may not be true. Depicted in blue are all observed targets, and in red are the EBs. The bottom panel is the flattened version of the previous panel, divided by the bottom envelope.}
    \label{fig:cdpp}
\end{figure}

\begin{figure}
    \centering
    \includegraphics[width=\textwidth]{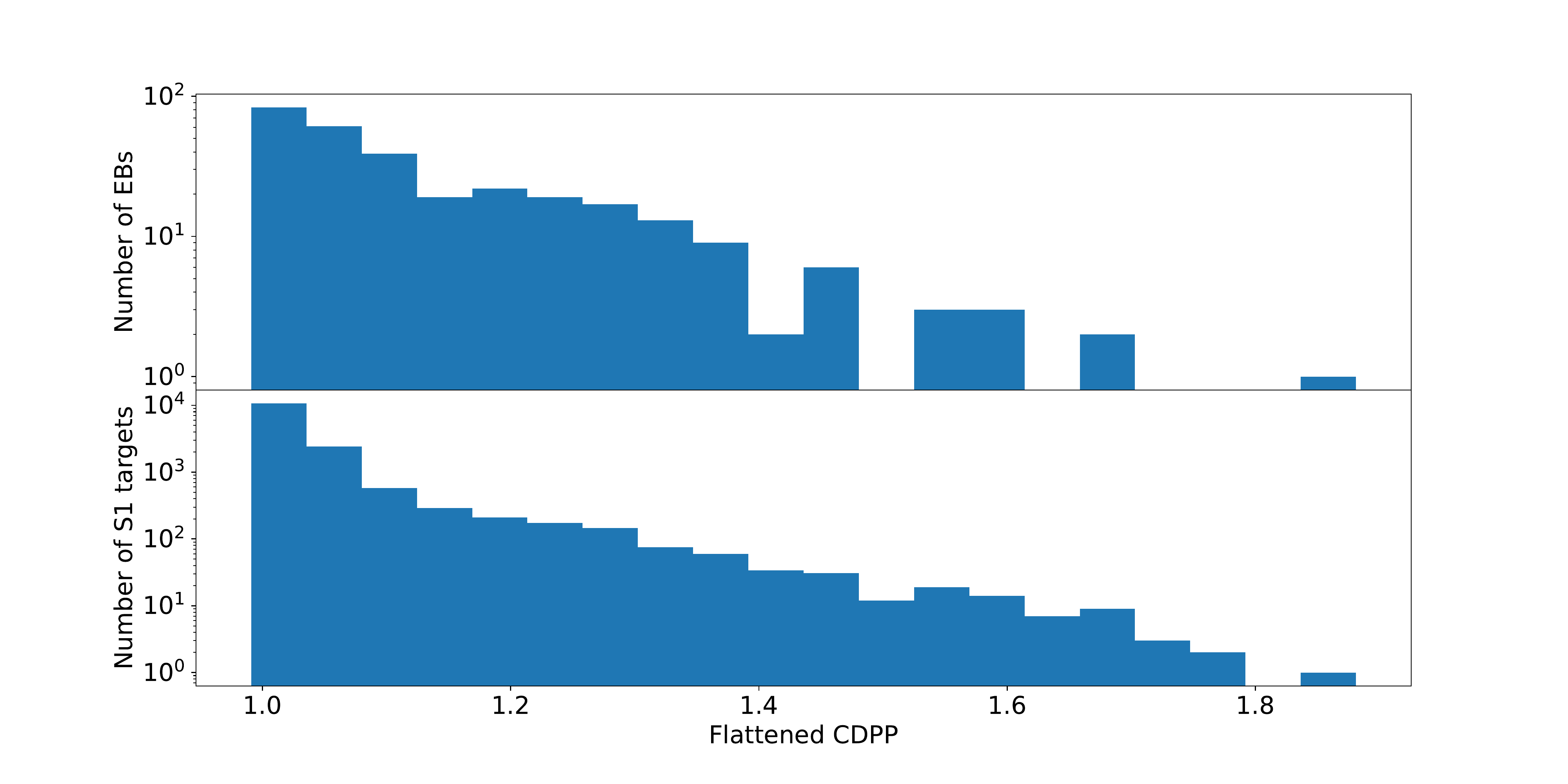} \\
    \caption{The number of EB targets (top) compared to all Sector 1 targets (bottom) as a function of flattened CDPP. Note the logarithmic $y$ scale.}
    \label{fig:completeness}
\end{figure}

To estimate completeness qualitatively, we start with the Combined Differential Photometric Precision (CDPP; \citealt{gilliland2011}). CDPP is a measure of light curve variation -- the rms noise of the result of filtering the timeseries with a high-pass Savitzky-Golay filter and then applying a moving average of 0.5-hr. The top panel in Fig.~\ref{fig:cdpp} shows CDPP as a function of magnitude for Sector 1 observations. The CDPP values are inherent to aperture light, so they are not corrected for crowding. The second panel in Fig.~\ref{fig:cdpp} provides a statistical estimate of per-target crowding, i.e., the fraction of light in the aperture coming from the target itself. Thus, a value of 1 means that all light is due to the observed target, while a value of, say, 0.6, means that 60\% of the light comes from the target and 40\% of the light comes from background sources. Thus, to account for dilution, we correct the CDPP for crowding by dividing it by the crowding factor. This implicitly assumes that the extraneous source of light in the aperture is constant, and that all variability comes from the target itself, which may or may not be true, but overall it plays a small role in correction because crowding is ``top-heavy'': 90\% of the targets have crowding factors larger than 0.95. The corrected CDPP is depicted in the 3rd panel in Fig.~\ref{fig:cdpp}. We then fit the bottom envelope by sigma-clipping and divide the corrected CDPP by the envelope in order to flatten out the dependence on magnitude (bottom panel). This way we can assess the distribution of amplitudes for EBs and its \emph{qualitative} resemblance to the expected geometric distribution (i.e., the probability of eclipses as a function of orbital elements). We emphasize the word ``qualitative'' here because, in order to calculate the true geometric distribution, we would need to be able to simulate target list selection, generate a synthetic sample of EBs and, from there, calculate the corresponding CDPP values; this, unfortunately, is not tractable because of the non-prescriptive target selection. The top panel in Fig.~\ref{fig:completeness} shows this distribution; we observe a monotonically decreasing trend, in line with expectations. For validation purposes we check the trend of all targets observed in Sector 1 -- as we see no systematic features in the flattened CDPP, we conclude that the number distribution is a fair reflection of the true CDPP distribution for the observed EBs.

While not statistically significant, we took a closer look at the slight dip in the 4th bin of the EB distribution; there are 19 EBs out of 299 targets in that bin that were found. We manually went through the entire list of 299 targets again but have failed to find any additional EBs.

\begin{figure}
    \centering
    \includegraphics[width=\textwidth]{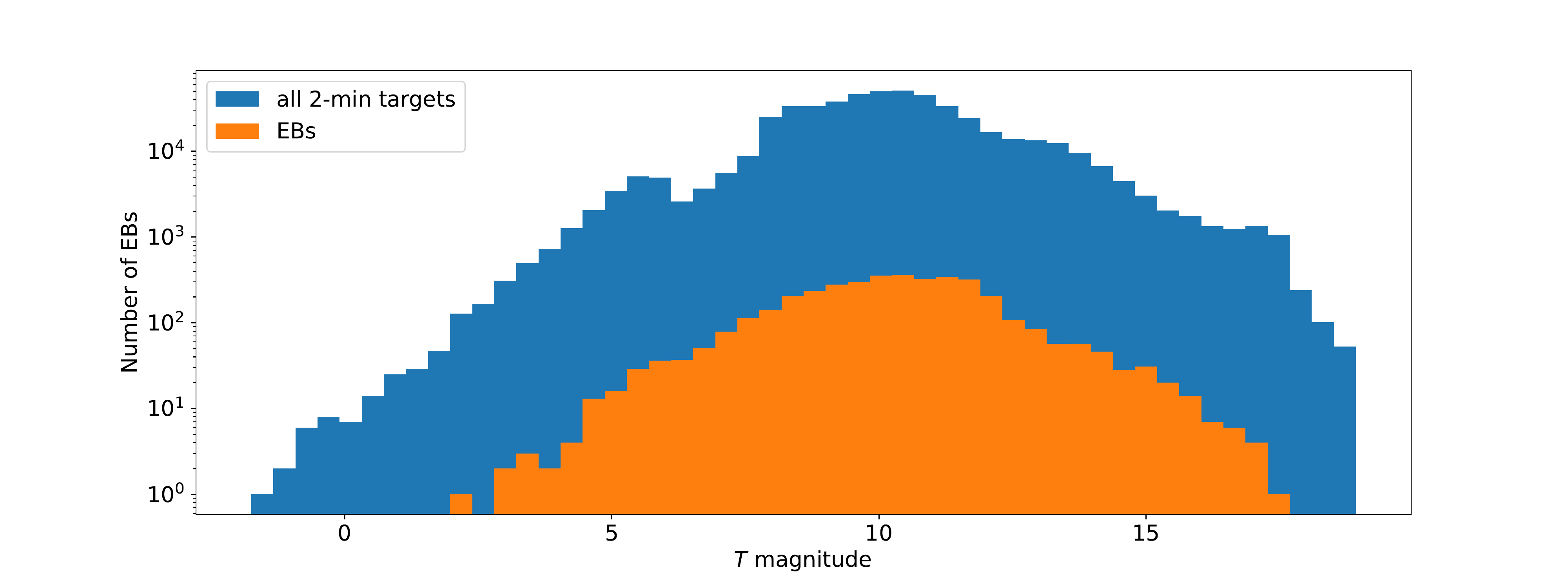} \\
    \caption{The distribution of \tess magnitudes for all 2-min targets observed in sectors 1--26 (blue) and all \ebno EBs (amber).}
    \label{fig:tmag}
\end{figure}

Finally, we perform one last check: we test whether the magnitude distribution of EB targets is representative of the overall magnitude distribution; any detected bias might raise completeness questions. Fig.~\ref{fig:tmag} demonstrates qualitative equivalence, noting of course that there are no EBs brighter than $T = 2.5$, and the diminishing signal-to-noise ratio suppresses eclipse detection on the faint end.

At some level eclipses will blend into the noise background and we will no longer be able to (reliably) detect them; to estimate that level, we take the distribution of primary eclipse depths (cf.~Fig.~\ref{fig:depths}) and choose the bin size that yields equal occurrence rates for the first two bins. Phenomenologically that implies that the monotonic increase towards lower SNR stops and that we can no longer reliably detect eclipses. That occurs at SNR$\sim$13. We take that to be our limit. We conclude that the catalog is largely complete to SNR$\sim$13.

\section{Contents of the catalog} \label{sec:catalog}

The \textsl{TESS} eclipsing binary catalog presented here contains all targets observed with a 2-min cadence in sectors 1 through 26 that are consistent with eclipsing binary signatures. The count as of this writing is \ebno EBs.

All data are available from two official servers: the project's main webpage, \url{http://tessEBs.villanova.edu}, and at MAST as a High Level Science Product via DOI \dataset[10.17909/t9-9gm4-fx30]{\doi{10.17909/t9-9gm4-fx30}}\footnote{\texttt{https://archive.stsci.edu/hlsp/tess-ebs}}. The official webpage hosts the ``rolling'' version of the catalog that is updated in real time as new targets are validated or existing targets are refuted. Once every quarter we take a snapshot of the rolling version, tag it with a release number and deliver it to MAST. Copies of the tagged releases are also kept on the project webpage.

Table \ref{tab:catalog} summarizes the database tables and columns contained in the catalog, along with the relationships between database tables. All entries in the \texttt{TIC} table are copied verbatim from the latest \textsl{TESS} input catalog (v8.1 at the time of this writing; \citealt{stassun2018}); all entries in the \texttt{Sector} table are taken from the latest version of the official pointing table\footnote{\texttt{https://tess.mit.edu/wp-content/uploads/orbit\_times\_20201013\_1338.csv}}; table \texttt{Origin} contains identifiers for persons or groups who contributed EB candidates (cf.~Section \ref{sec:detection}). Table \texttt{EB} is the central table of the database: it contains links to the \texttt{TIC}, all \texttt{Sector}s the EB is observed in, and all \texttt{Origin}s. In addition, it provides the data source\footnote{In this work we only account for the 2-min cadence LCF observations, but the database allows for adding FFI and TPF data as they are ingested into the catalog.} (FFI for full-frame images; TPF for target pixel files; and LCF from lightcurve files). It also gives the refined ephemerides (cf.\ Section \ref{sec:refinement}) and morphology coefficient (cf.\ Section \ref{sec:morph}). Table \texttt{EphemerisSource} aggregates all automated ephemeris finding algorithms (cf.\ Section \ref{sec:ephems}) and table \texttt{Ephemeris} keeps all records on the proposed ephemerides. Note that these are \emph{not} final, refined ephemerides; rather, these are ``raw'' ephemerides output by the ephemeris finding algorithms. The \texttt{Ephemeris} table also stores triage information: when the triage was done, by whom, what was the disposition, and what proposed ephemerides are likely correct. Finally, the \texttt{Comment} table contains all subjective comments made on either \texttt{Ephemeris}, \texttt{EB} or \texttt{TIC} entries.

\startlongtable
\begin{longrotatetable}
\begin{deluxetable}{lllll}
\tablecaption{Database structure of the EB catalog. \label{tab:catalog}}
\tablenum{1}

\tablehead{
    Table & Column & Units & Value type & Description \\
}

\startdata
\textbf{TIC:}
    & \texttt{tess\_id} & ---      & long integer    & TIC identifier \\
    & \texttt{ra}       & deg      & float [0, 360)  & right ascension \\
    & \texttt{dec}      & deg      & float [-90, 90] & declination in deg \\
    & \texttt{glon}     & deg      & float [0, 360)  & Galactic longitude \\
    & \texttt{glat}     & deg      & float [-90, 90] & Galactic latitude \\
    & \texttt{Tmag}     & ---      & float & \textsl{TESS} magnitude \\
    & \texttt{teff}     & K        & float & estimated effective temperature \\
    & \texttt{logg}     & log(cgs) & float & estimated surface gravity \\
    & \texttt{abun}     & log(sol) & float & estimated metal abundances \\
    & \texttt{pmra}     & mas/yr   & float & proper motion in right ascension \\
    & \texttt{pmdec}    & mas/yr   & float & proper motion in declination \\
\textbf{Sector:}
    & \texttt{sector\_id}       & --- & integer         & Sector identifier \\
    & \texttt{date\_start}      & iso & date            & start date of sector observations \\
    & \texttt{date\_end}        & iso & date            & end date of sector observations \\
    & \texttt{spacecraft\_ra}   & deg & float [0, 360)  & satellite right ascension \\
    & \texttt{spacecraft\_dec}  & deg & float [-90, 90] & satellite declination \\
    & \texttt{spacecraft\_roll} & deg & float 0, 360]   & satellite roll from reference point \\
    & \texttt{cameraN\_ra}      & deg & float [0, 360)  & camera \texttt{N} right ascension, $\texttt{N}=1 \dots 4$ \\
    & \texttt{cameraN\_dec}     & deg & float [-90, 90] & camera \texttt{N} declination, $\texttt{N}=1 \dots 4$ \\
    & \texttt{cameraN\_roll}    & deg & float 0, 360]   & camera \texttt{N} roll from reference point, $\texttt{N}=1 \dots 4$ \\ 
\textbf{Origin:}
    & \texttt{name} & --- & string & Data/classification origin string \\
\textbf{EB:}
    & \texttt{tic}        & --- & \texttt{TIC}    & link to the TIC entry \\
    & \texttt{origin}     & --- & \texttt{Origin} & link to the origin entry \\
    & \texttt{sectors}    & --- & \texttt{Sector} & list of sectors the EB was observed in \\
    & \texttt{signal\_id}   & --- & integer         & EB signal enumerator \\
    & \texttt{in\_catalog} & --- & boolean         & flags the EB as validated \\
    & \texttt{date\_added} & iso & date            & time and date of the database entry \\
    & \texttt{date\_modified} & iso & date            & time and date of the database modification \\
    & \texttt{source}         & --- & choice       & data source: 'FFI', 'TPF' or 'LCF' \\
    & \texttt{bjd0}           & days & float & BJD$-2457000$ of the primary eclipse \\
    & \texttt{bjd0\_uncert}  & days & float & uncertainty in bjd0 \\
    & \texttt{period} & days & float & orbital period \\
    & \texttt{period\_uncert}  & days & float & uncertainty in orbital period \\
    & \texttt{morph\_coeff} & --- & float & morphology coefficient \\
    & \texttt{morph\_dist}  & --- & float & orthogonal distance from the morphology manifold \\
\textbf{EphemerisSource:}
    & \texttt{model}     & --- & string & Model used for the ephemeris estimate \\
    & \texttt{version}   & --- & string & Model version \\
    & \texttt{author}    & --- & string & Model author \\
    & \texttt{reference} & --- & string & Model reference \\
\textbf{Ephemeris:}
    & \texttt{date\_added} & iso & date & Ephemeris estimate date \\
    & \texttt{source} & --- & \texttt{EphemerisSource} & Ephemeris source \\
    & \texttt{eb} & --- & \texttt{EB} & link to the EB entry \\
    & \texttt{bjd0}           & days & float & BJD$-2457000$ of the primary eclipse \\
    & \texttt{bjd0\_uncert}  & days & float & uncertainty in bjd0 \\
    & \texttt{period} & days & float & orbital period \\
    & \texttt{period\_uncert}  & days & float & uncertainty in orbital period \\
    & \texttt{triage\_timestamp} & iso & date & time of performed triage \\
    & \texttt{triage\_status} & --- & string & triage disposition \\
    & \texttt{triage\_period} & --- & choice & choice of 'period', 'half', 'double', 'other' \\
    & \texttt{triage\_username} & --- & string & person completing the triage \\
\textbf{Comment:}
    & \texttt{author} & --- & string & comment author \\
    & \texttt{text} & --- & string & comment text \\
    & \texttt{timestamp} & iso & date & comment date and time \\
    & \texttt{ephem} & --- & \texttt{Ephemeris} & link to the ephemeris entry \\
    & \texttt{eb} & --- & \texttt{EB} & link to the EB entry \\
    & \texttt{tic} & --- & \texttt{TIC} & link to the TIC entry \\
\enddata

\end{deluxetable}
\end{longrotatetable}

\section{Discussion} \label{sec:discussion}

In addition to curating information on individual EB targets observed by \tessnts, the catalog can be used for bulk analysis, similar to what was presented in Section \ref{sec:props}. Here we present another two interesting observations.

\begin{figure}
    \centering
    \includegraphics[width=\textwidth]{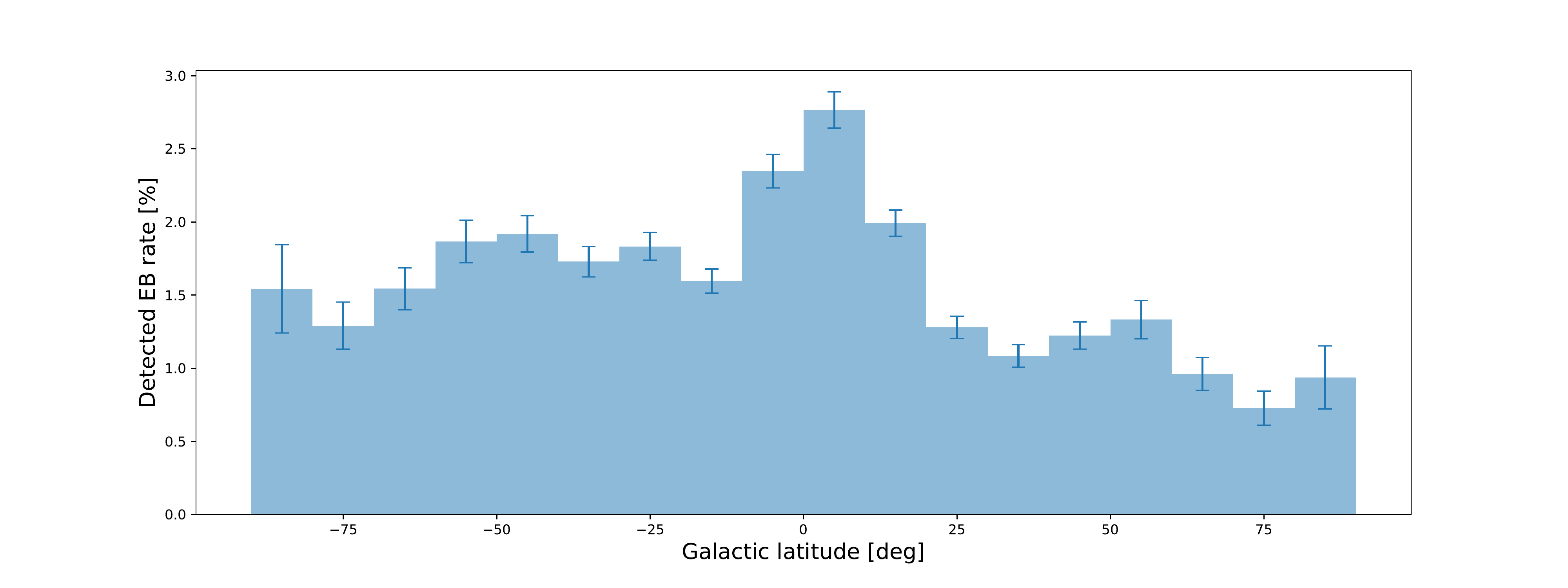} \\
    \caption{Detection frequency of EBs as a function of Galactic latitude. The frequency is computed by counting all detected 2-min cadence EBs in a given 10-degree Galactic latitude strip and dividing the count with the number of all 2-min cadence targets observed in the same strip.}
    \label{fig:glatdist}
\end{figure}

Fig.~\ref{fig:glatdist} depicts galactic latitude distribution of EBs, with two notable features. The first is a near-constant distribution of southern galactic hemisphere EBs ($-90 < b < -10$), and the second is the lopsidedness between the southern and the northern galactic hemisphere. The near-constant distribution of EBs is at odds with observations from, say, \kepler \citep{prsa2011}, where galactic latitude dependence was pronounced. This is a consequence of (1) the non-uniform target selection presented in Section \ref{sec:detection}; (2) the relatively bright magnitude breakdown limit for \tessnts, significantly reducing the dependence of eclipse probability on stellar populations; and (3) the fraction of time the Galactic plane was in or near the CVZ. We expect that EBs detected in FFIs will comprise a more uniform sample that will be more representative of true galactic distributions of EBs (cf.~Fig.~\ref{fig:ffi_ebcat_coords}). The lopsidedness is largely due to the change in boresight in sectors 14-16 and 24-26 (cf.~Fig.~\ref{fig:tess_ebs}).

\begin{figure}
    \centering
    \includegraphics[width=\textwidth]{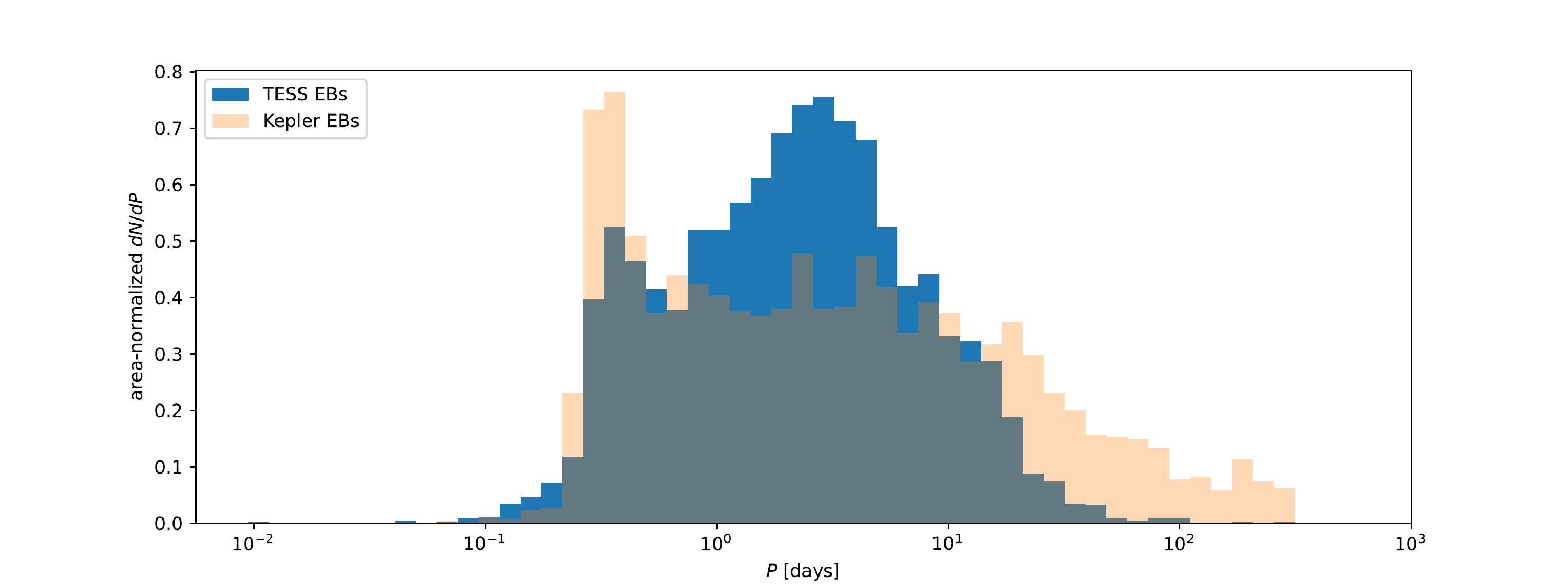} \\
    \includegraphics[width=\textwidth]{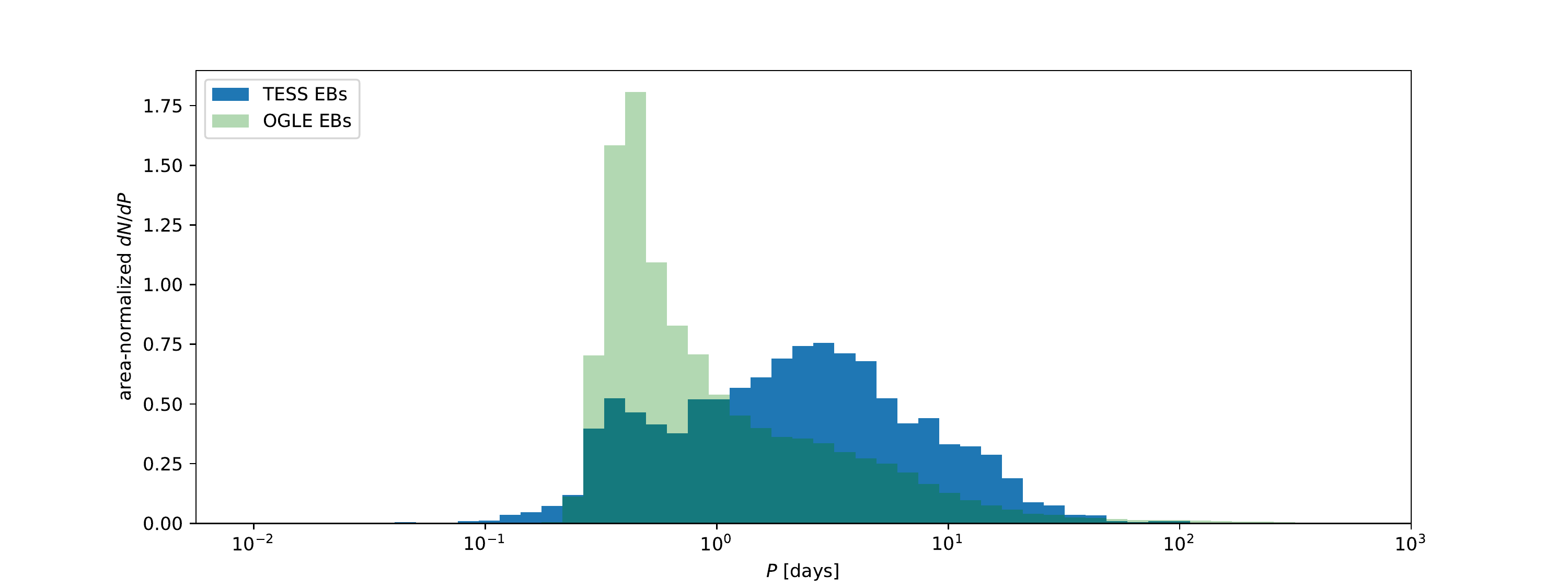} \\
    \caption{The comparison of orbital period distributions from \tessnts, \kepler and OGLE. All distributions are normalized by their surface-areas for comparison purposes.}
    \label{fig:logpcomp}
\end{figure}

The second observation is bias in orbital periods of \tess EBs. Fig.~\ref{fig:logpcomp} (top) shows a comparison between the orbital period distributions for \tess EBs (blue) and \kepler EBs (amber). The long tail for the \kepler sample on the long period end is expected given that \kepler observed the same field for $\sim$4 years; on the short period end we see concordance of the narrow peak around $P\sim 0.4$ days between the two datasets, corresponding to contact binary stars, but a different distribution of EBs with periods around 5 days: \keplernts's distribution is mostly flat, while \tess distribution features a wide peak. This is again a consequence of temporal coverage: the 5-day orbital periods are the sweet spot for \tess with $\sim$5 cycles per observed sector, compared to $\sim$300 cycles for \kepler. Thus, \keplernts's extended sensitivity towards longer periods means that its detection rates are suppressed almost exclusively by the diminishing geometric probability of eclipses; in contrast, \tess suffers from duty cycle suppression at orbital periods as short as $\sim$13 days, where geometric suppression is not yet dominant.
Fig.~\ref{fig:logpcomp} (bottom) depicts a comparison of \tess EBs against the ground-based survey OGLE (green). In this case we see that the contact binary peak in the OGLE dataset is severely enhanced by the diurnal selection effect. Longer periods are significantly suppressed both by the duty cycle limitations, as well as geometric probability of eclipses that is now more pronounced because of the sparser, irregular observing cadence. This bias clearly highlights the benefits of space-based surveys such as \tessnts.

\subsection{Goals of the paper series} \label{sec:goals}

In this first part of the series of papers we presented the sample of \ebno EBs observed by \tess in sectors 1--26. The main science goals that the entire series aims to achieve are as follows:
\begin{description}
\item[Study gold standard EBs] These are known, bright, detached, double-lined binaries with deep eclipses \citep{torres2010, southworth2015}. These are posited to have had minimal interaction histories and have evolved largely independently, as single stars. Such systems are considered ``gold standards'' because they hold promise for the most accurate fundamental parameters. These EBs have spectroscopic data already available, and \tess provides us with an improvement in the photometric light curve quality and temporal coverage.

\item[Estimate photometric elements for all \tess EBs]

Eclipsing binary light curves hold a wealth of information on the system and its stellar components. Using \textsc{PHOEBE} and its framework for inverse problem solvers \citep{Conroy2020}, we can estimate photometric parameters of all \tess EBs with minimal manual intervention. Using the ephemerides and geometrical estimates from the catalog as a starting point, we can determine the sum and ratio of fractional radii, temperature ratio of the components; the inclination, eccentricity and argument of periastron of the orbit; as well as passband luminosities and potential third light contribution. In addition, for close systems, the mass ratio and semi-major axis \rev{can be constrained from light curves alone \citep{wilson1994,terrell2005}}. A full Bayesian analysis with MCMC also yields posterior distributions for second-order effects, like the effective temperature of the primary, gravity darkening coefficients and albedoes, which, given the photometric precision of \tessnts, can be used to refine the current empirical laws.

The main perceived impact of this study will be the period-eccentricity distribution of the \tess EBs sample, which is crucial in studies of binary star populations. In addition, interesting and benchmark systems will be identified for radial velocity follow-up and more advanced analysis.

\item[Extend the temporal coverage with archival photometry]

TESS observations last about 27 days for each sector.  Roughly 70\% of the sky was observed during the prime mission \citep{guerrero2021}, and of the 232,705 stars observed at 2-minute cadence, 152,993 ($\sim66$\%) were observed for just one sector, while the rest were observed in multiple sectors, and 3874 ($\sim$1.7\%) which were observed for 13 consecutive sectors, over 351 days.  While the observations are not fully continuous, with gaps between orbits and some time lost for engineering or safe modes, in general a given time baseline of observations means that it is possible to detect EBs with orbital periods up to half the time baseline of observations.

At the long end of that detectability time frame, ambiguities exist in the characterization of the EBs.  In some cases, it is not clear whether two distinct eclipses are seen.  In others, spot modulation or other variability can interfere with the identification or measurement of an eclipse.  For EBs with orbital periods greater than half the observing time baseline, only a single eclipse may be detected, yielding a likely identification of an EB without a reliable period measurement.  See examples in Figure \ref{fig:single-transits}.  For all these reasons, the set of TESS-detected EBs can benefit from cross-identification with other photometric surveys.

\begin{figure}
    \centering
    \includegraphics[width=\textwidth]{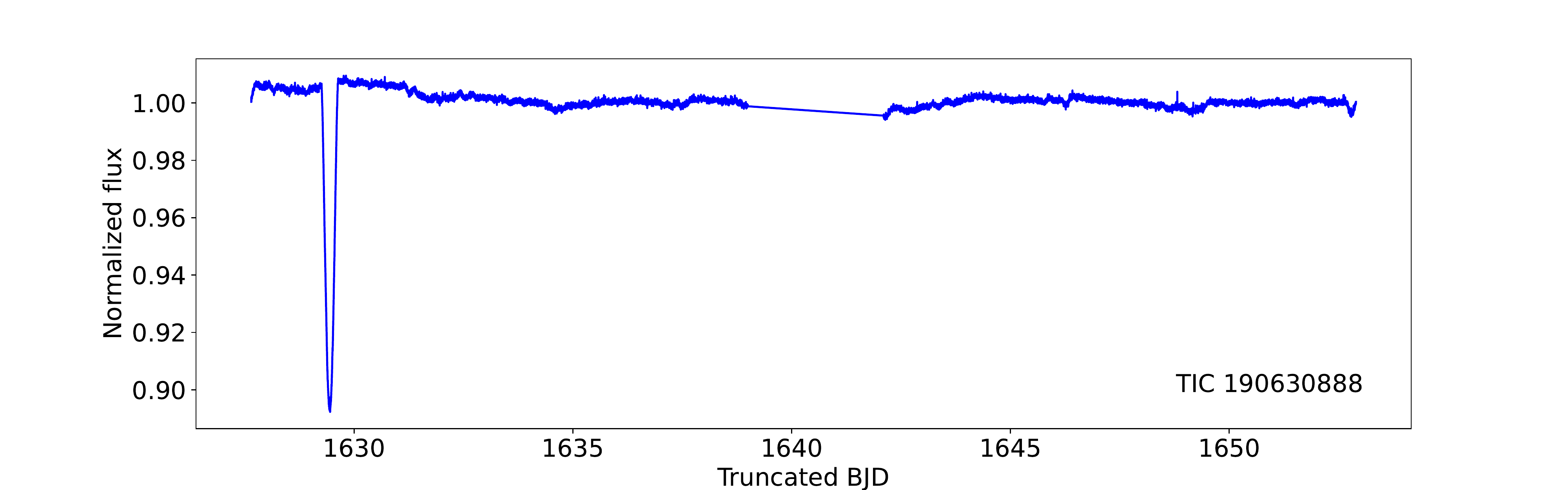} \\
    \includegraphics[width=\textwidth]{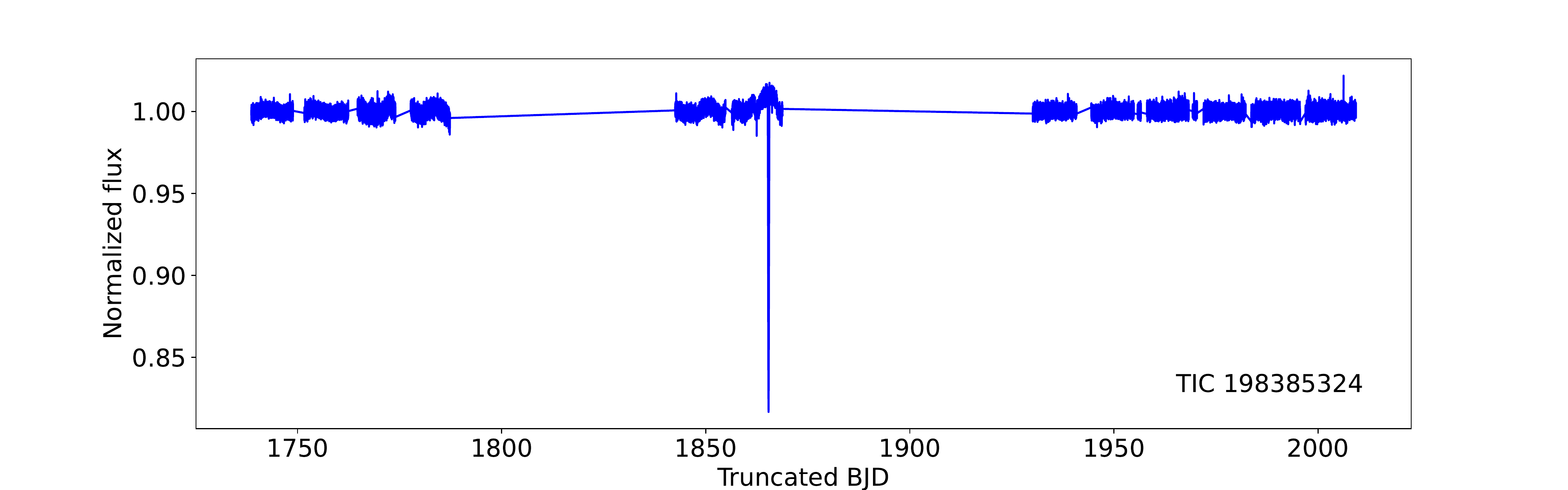} \\
    \caption{Examples of single-transit EBs, TIC 190630888 and TIC 198385324.}
    \label{fig:single-transits}
\end{figure}

A number of such surveys have been conducted, and can be combined with TESS observations.  Many surveys were carried out to search for transiting exoplanets, including WASP \citep{Pollacco:2006}, HAT \citep{Bakos:2004,Bakos:2013}, KELT \citep{Pepper:2007,Pepper:2012}, and others \footnote{See \citep{Deeg:2018} for a larger list}.  Other surveys have been conducted to search for stellar variability, such as ASAS \citep{Pojmanski:1997}, ASAS-SN \citep{Shappee:2014}, ATLAS \citep{tonry2018}, ZTF \citep{Bellm:2019}, etc.  While none of these surveys have the photometric precision or cadence of TESS, they all have greater observing time baselines over large fractions of the sky.  EBs identified in TESS data with ambiguous ephemerides can be matched to the archival photometric observations to identify eclipses and refine the ephemerides.  Such an effort will typically only recover EBs with deep eclipses, but that still represents a large fraction of the EB population.  The archival photometry can determine or refine the ephemerides of the EBs, and can be used to search for eclipse time variations (ETVs; \citealt{conroy2014}) or apsidal motion [REF]\citep{Borkovits16,Orosz15,Kirk16}.  Further development of the TESS EB catalog will incorporate such analysis.

In addition to the use of archival photometry, individual systems of interest can be followed up for astrophysical investigation with modest observing facilities.  While broader scientific themes are outlined below, small-telescope observatories are well-equipped to observe the bright EBs identified in this catalog.  An example of that kind of investigation is multiband photometric follow-up of the O'Connell effect \citep{Milone:1968}, in which light curves show unequal maxima before and after the primary eclipse.  Follow-up observations in different bandpasses are necessary to understand the underlying cause of this phenomenon. Furthermore, while archival photometry can recover the orbital period for some long-period EBs, as described above, follow-up photometry can readily refine the orbital parameters more precisely.

\item[Study asteroseismic components in EBs] 

The exquisite precision, cadence, and duty-cycle of TESS observations have revealed resolved pulsations in stars across all spectral types and evolutionary phases \citep{Pedersen2019,Antoci2019,Corsico2021}. Given the ubiquity of both stellar pulsations and stellar multiplicity, binary systems with at least one pulsating component are commonly found. The complementary nature of asteroseismic (interior) and binary (bulk) information enables highly detailed evolutionary modelling. Such synergistic combined modelling approaches have led to detailed inference on the structure of stars from solar-like oscillators to massive $\beta$-Cep pulsators to evolved stars which have undergone mass transfer \citep{deCat2004,Schmid2016,Guo:2016,White2017,Guo2017,Beck2018,Streamer:2018,Johnston2019,johnston:2021}. Furthermore, detailed studies have demonstrated tension between the asteroseismic and binary results for both main sequence and evolved stars, indicating that evolutionary models still have room for improvement \citep{gaulme2016,Themessel:2018,Benbakoura:2021,Sekaran:2021}. The large data-set provided by TESS will enable synergistic asteroseismic and binary modelling across a much larger parameter range than was previously possible with the more limited samples, allowing investigation into the relationship between pulsational properties and orbital properties \citep{Liakos:2017,Gaulme:2019,Sekaran:2020}.

Ground-based radial velocity follow-up of binary systems is time consuming, resource intensive, and simply expensive. Due to their stability over long time bases, heat-driven pulsations act as regular clocks, whose phase (and frequency) is modulated due to the light travel-time effect from the orbital motion. By measuring the observed phases of pulsations across the orbit, one can reliably derive the radial velocity curve of the pulsating component from the photometry alone \citep{Shibahashi2012,Murphy2014,Hey2020}. The application of this method to this database (where appropriate given the ratio of pulsation period to orbital period) can produce a wealth of radial velocity curves without needing to apply for costly ground-based telescope time. 

\item[Explore the impact of dynamical tides]

Kepler discovered 172 heartbeat stars (HBs), and about one fifth show tidally excited oscillations \citep{kirk2016,Guo2020}, i.e., the direct manifestation of dynamical tides. A handful of them have been studied in detail \citep{welsh2011,Hambleton2013,Hambleton2016,Hambleton2018,Shporer2016,Fuller2017,Guo2017,Guo2019}. Recently, Guo (2021) reviewed the current status of heartbeat binaries with tidally excited oscillations. Most Kepler HBs are of A--F types due to selection biases. TESS is destined to discover more massive O--B type heartbeat binaries. Systems already published include \citet{Jayasinghe2019,Jayasinghe2021} and \citet{Kolaczek-Szymanski2021}. To date, we have identified 25 HB systems in the 2 minute TESS data up to Sector 26, and over 200 systems in the 2 minute and FFI data, combined. Given that 5\% of the binary stars in the \kepler\ catalog with periods less than 27.3\,d (the length of a \tess\ sector) were classified as heartbeat stars, we anticipate that the number we have identified is a lower limit.    

The advent of the TESS mission has also revealed a new class of pulsating stars in binaries, wherein the pulsation axis is ``tidally tilted'', producing a characteristic amplitude and phase modulation \citep{Handler2020,Kurtz2020,Fuller2020,Rappaport2021}. Additionally, a few studies have reported the detection of so-called ``tidally perturbed'' pulsations, where a series of pulsations that are nearly equidistantly spaced by the orbital frequency were detected \citep{Bowman2019,Steindl2021,Southworth2020}. Such systems cannot currently be explained by the same model as the ``tidally tilted'' pulsators, and seem to have a separate connection to the interplay between tides and pulsations.
While only a handful of these objects have been identified to date, they signify a unique contribution of TESS to the progress of tidal asteroseismology. The compilation of this database serves to streamline the detection of even more systems.

\item[Stellar multiples and circumbinary planets]


A large number of eclipsing binaries have third body companions that are detected via eclipse timing variations (ETVs), either light travel time effects (LTTE, the most common) or dynamical effect delays. A fraction of these exhibit tertiary eclipses as well.  For example, \citet{Borkovits2016} found good evidence for third bodies in 222 eclipsing binaries out of 2600 studied using {\em Kepler} data.  In most cases, the ETV signal becomes significant on time scales similar to or longer than the period of the outer orbit, which is typically a few to several hundred days for the systems found by \citet{Borkovits2016}.  

Rather than being a star, in some cases the third body is a planet - a so-called circumbinary planet. Kepler revealed 13 such planets that transited their host stars (e.g.\ see \citet{Doyle2011}, \citet{Welsh2012}, \citet{Orosz2019}). While few in number, these systems provide a great deal of information, e.g., extremely accurate (not just precise) stellar parameters and planetary radii. But with only a handful of systems we cannot understand the characteristics of the population of circumbinary planets. TESS's all-sky survey could remedy this small sample-size problem, but for most of the sky only $\sim$27 days of near-continuous coverage per two years is available and this window is much shorter than the typical circumbinary planet orbital period. Fortunately the duty cycle at the ecliptic poles is much better and the first TESS circumbinary planet, TOI-1338 \citep{Kostov2020a} was detected in observation from TESS's continuous viewing zone (CVZ) at the southern ecliptic cap. 
But in addition to the long-duration observations at the CVZ, we can search for circumbinary planets by taking advantage of the extra information provided when the planet transits both star in the binary. Unlike a planet orbiting a single star, a circumbinary planet can produce two, and possibly more, transits during one conjunctions pass. This ``1-2 punch'' technique \citep{jenkins1996,Kostov2020b} was used to detect the second TESS circumbinary planet, TIC 172900988 \citep{Kostov2021}. Roughly 140 $\pm 110$ new circumbinary planets may potentially be found via this technique, in addition to the 40 $\pm$ 32 expected in the two CVZs \citep{Kostov2020b}. Thus TESS should be able to boost the number of circumbinary planets by an order of magnitude, allowing a much more robust estimate of the occurrence rate, population statistics, and searches for correlation with various binary star parameters (e.g. metallicity, eccentricity, mass ratio, etc.).

\item[Gaia data overlap]

Of the \ebno EBs in the catalog, a total of 3486 have entries in the early third data release of the Gaia catalog \citep[Gaia EDR3;][]{Gaia:2021}. The top three panels in Figure~\ref{fig:gaiahist} show histograms of the $G$-band magnitudes and the $G_{\rm BP}-G_{\rm RP}$ colors for the combined light of each binary, and of their distances, where the latter is derived here simply as the reciprocal of the trigonometric parallax \citep[see][]{Bailer-Jones:2021}. The distribution of magnitudes up to about $G = 12$ largely reflects the content of the TIC Candidate Target List, while the fainter stars are special interest targets that come from the Guest Investigator Program. The typical color index is very near solar. As expected from their brightness, most of these objects are relatively nearby, with the peak of the distribution located at $\sim$150~pc, and a long tail toward larger distances (truncated in the figure).

\begin{figure}
\plotone{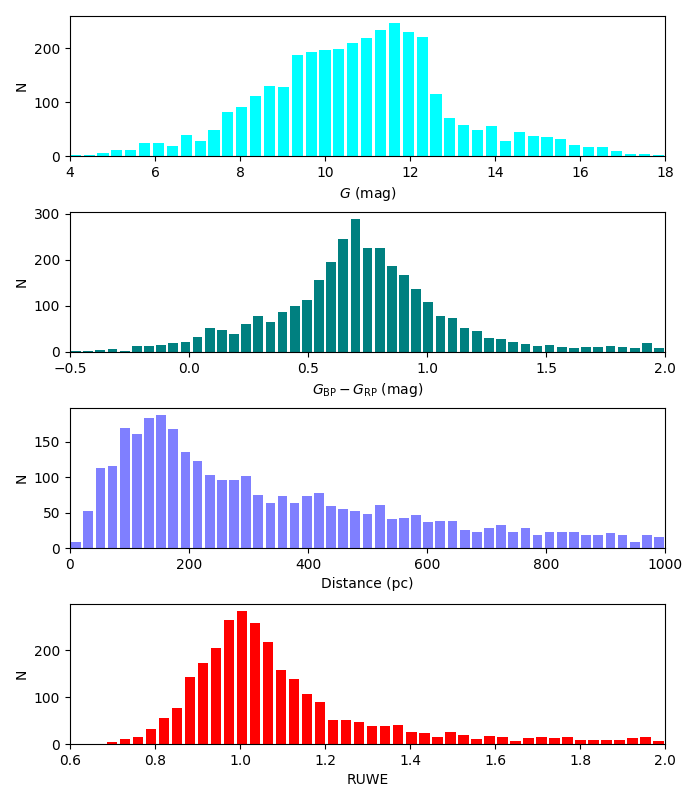}
\figcaption{Distributions of the Gaia EDR3 brightness, color, distance, and RUWE for the eclipsing binaries in the catalog. The histograms are truncated on the right (and some on the left) for clarity.\label{fig:gaiahist}}
\end{figure}

The bottom panel displays the distribution of the Renormalized Unit Weight Error (RUWE) from Gaia, which is regarded as a useful measure of the quality and reliability of the astrometric solution\footnote{See report by L.\ Lindegren in the Gaia documentation, \url{https://www.cosmos.esa.int/web/gaia/public-dpac-documents}}. RUWE values for well-behaved solutions cluster around 1.0, whereas values larger than about 1.4 typically indicate unmodeled excess scatter in the astrometric observations that can be caused, e.g., by binarity of the source, or perhaps other effects. Gaia does not spatially resolve any of the EBs in this catalog, it only measures the center of light. In many cases the binary motion will cause the flux centroid to wobble enough to be detected by Gaia, perturbing the astrometric solution.

\cite{Stassun:2021} have shown that even in the 1.0--1.4 range the RUWE values tend to correlate with the semimajor axis of the photocentric motion. For the binaries in the catalog we do not have sufficient information to estimate the semimajor axis of the photocenter. However, we can place the EB sample into the broader context of typical RUWE values for Gaia stars across the HR diagram. This is shown in Figure~\ref{fig:gaiacmd1}, where the left panel represents the RUWE in grayscale for the $\sim$1~million Gaia EDR3 stars within 400~pc having $5 < G < 19$ and $|b| > 60^\circ$, and the right panel overlays in green the EB sample. In both panels, a piecewise linear fit is shown over the ``spine" of the single-star main sequence, for which the RUWE values are at a minimum (dark gray). Various populations with relatively high binary-star fractions appear above and below the single-star main sequence (lighter shades of gray), as described in detail by \citet{Belokurov:2020}. 

\begin{figure}[!ht]
    \centering
    \includegraphics[width=0.49\linewidth,trim=0 0 0 50,clip]{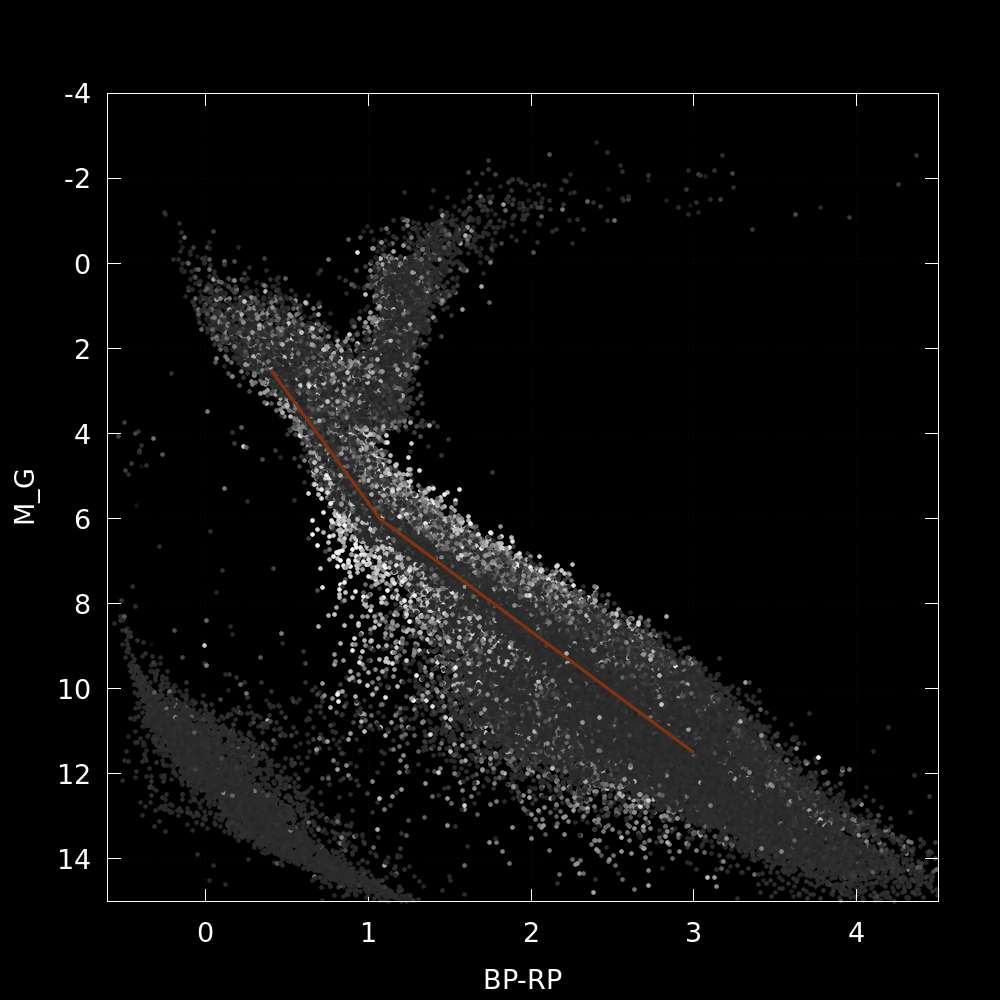}
    \includegraphics[width=0.49\linewidth,trim=0 0 0 50,clip]{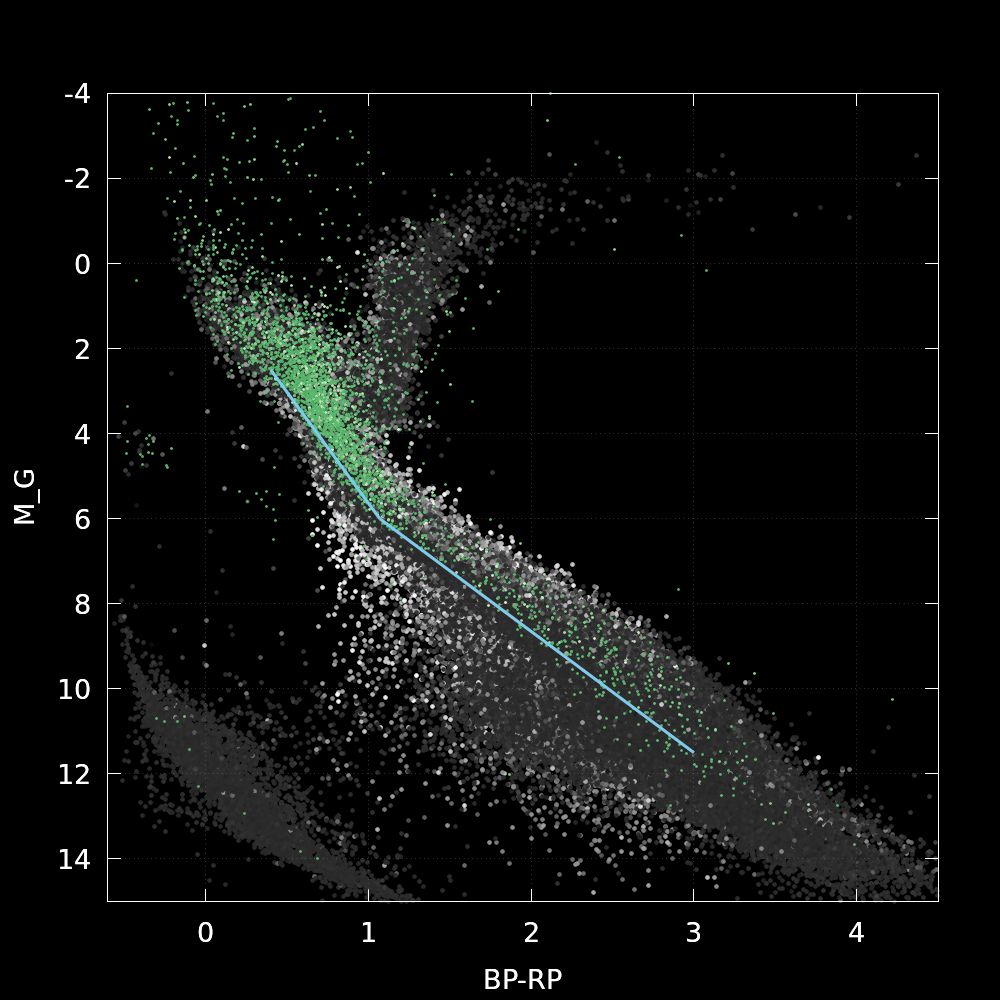}
    \caption{The color-magnitude diagram in the Gaia passbands. The grayscale denotes the Renormalized Unit Weight Error (RUWE) parameter for $\sim$1 million Gaia EDR3 targets within $\sim$400\,pc. Left: Gaia-only targets, with the piecewise linear line is the fit to the lowest RUWE values. Right: \tess EBs plotted over Gaia targets; most EBs are displaced to the brighter, redder region of the CMD.}
    \label{fig:gaiacmd1}
\end{figure}

Relative to the single-star main sequence, the EBs identified in this paper are in almost all cases displaced upward/redward, as expected for the combined light of pairs of stars having various relative luminosities and colors. And relative to the minimal RUWE values typical of the single-star main sequence, the EBs occupy regions that are characterized by higher RUWE values on average. 

We can also use the recent work of \citet{Belokurov:2020} to assess the likely nature and evolutionary states of the EBs. In Figure~\ref{fig:gaiacmd2}, we represent the EB sample in the color-magnitude diagram with theoretical evolutionary tracks overlaid to provide a sense of the mass range of the systems, as well as boxes with annotations indicating EBs in evolutionary states that may be of particular interest for various astrophysical applications. 

\begin{figure}[!ht]
    \centering
    \includegraphics[width=0.7\linewidth]{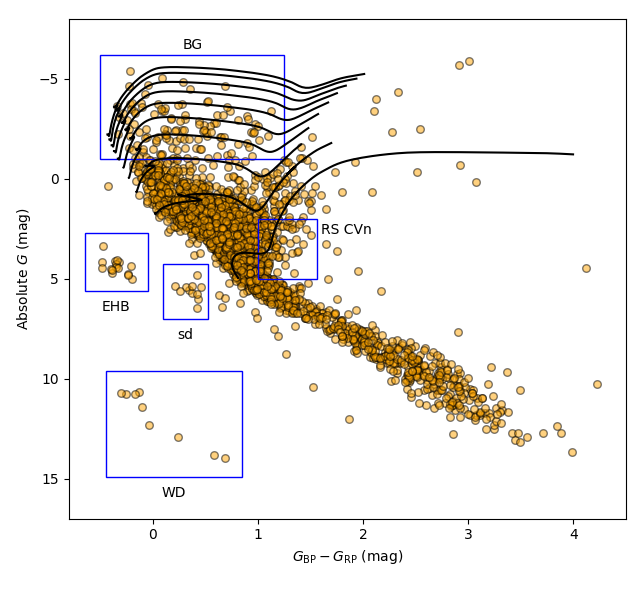}
    \caption{EB sample in the color-magnitude diagram with theoretical evolutionary tracks spanning 1--10 solar masses, and boxes indicating evolutionary states of particular interest \citep[see, e.g.,][]{Belokurov:2020}, including white dwarfs (WD), hot subdwarfs (sd), extreme horizontal branch (EHB) stars, RS CVn systems, and a remarkably large number of bright giants (BG; i.e., giant stars of intermediate mass).}
    \label{fig:gaiacmd2}
\end{figure}

Finally, it is now well established that short-period binary stars are very frequently found in hierarchical triple systems, with a tertiary companion on a much larger orbit around the tight binary \citep[see, e.g.,][and references therein]{Tokovinin:2006, Laos:2020}. We cross-matched our EB sample with the catalog of wide binaries in Gaia EDR3 from \citet{El-Badry:2021}, to assess if any of the EBs are in fact members of hierarchical triples.  Curiously, no matches were found.
\rev{Any tertiaries separated from the EB by less than 1\farcs5 will likely be unresolved and missing in EDR3, and those with separations less than about 0\farcs5 will almost certainly be unresolved; hence, it may not be surprising that \citet{El-Badry:2021} did not report resolved tertiaries for these systems. More importantly, we re-emphasize that the EB sample clearly exhibits systematically elevated RUWE values in Gaia (Figure~\ref{fig:gaiacmd1}), which is likely to be a direct manifestation of unresolved tertiaries in many cases, especially for those EBs with $P<3$~d \citep[see][]{Tokovinin:2006}.}

\item[Use calibrated 2-min cadence targets for FFI extraction] There are several community-led projects that focus on lightcurve extraction from \tess FFIs. The tools and the data products stemming from those projects have been released to the public; we base our extractions on an adapted version of \texttt{eleanor} \citep{feinstein2019}. We use the EBs observed with the 2-min cadence as the training set for the backpropagating neural network that will sift through all FFI lightcurves and find those that resemble EBs. The amount of cyan in Fig.~\ref{fig:tess_ebs} clearly showcases why manual vetting is \emph{not} tractable and why we \emph{must} resort to automated techniques. We will extract FFI lightcurves for each of the 2-min cadence EBs, formulate a fidelity metric based on the comparison between FFI-extracted and 2-min cadence lightcurves that can be inspected manually, train a classifying, back-propagating neural network on the FFI-extracted lightcurves of the 2-min cadence EBs, and deploy the network on \emph{all} FFI lightcurves to automatically detect EBs lurking in the images.

\item[Analyze contents of a catalog of EBs extracted from the FFIs] Several of our co-authors have constructed all light curves from {\em TESS} FFIs up to 15th magnitude, resulting in more than 72 million light curves from sectors 1-26.  The authors constructed a one-dimensional convolutional neural network, first described in \citet{2021AJ....161..162P}, to extract candidate EBs from these light curves, resulting in a catalogue of over 460,000 sources, of which we expect ~250,000 to be true sources (i.e. not due to light curve contamination).  The coordinates of these candidates are shown in Figure \ref{fig:ffi_ebcat_coords}.  Among these candidates were found several interesting multiple star systems \citep{2021AJ....161..162P,2021ApJ...917...93K}, with many others currently being researched.  The greatest value of this effort, however, may indeed lie in the use of the full catalog of candidates.  Cross-matching with other catalogs could identify higher order multiples or binaries containing particular stellar types.  Statistical analysis of the catalog could also provide tremendous insight to the growing field of EB research. 

\begin{figure}[!ht]
    \centering
    \includegraphics[width=1.0\linewidth]{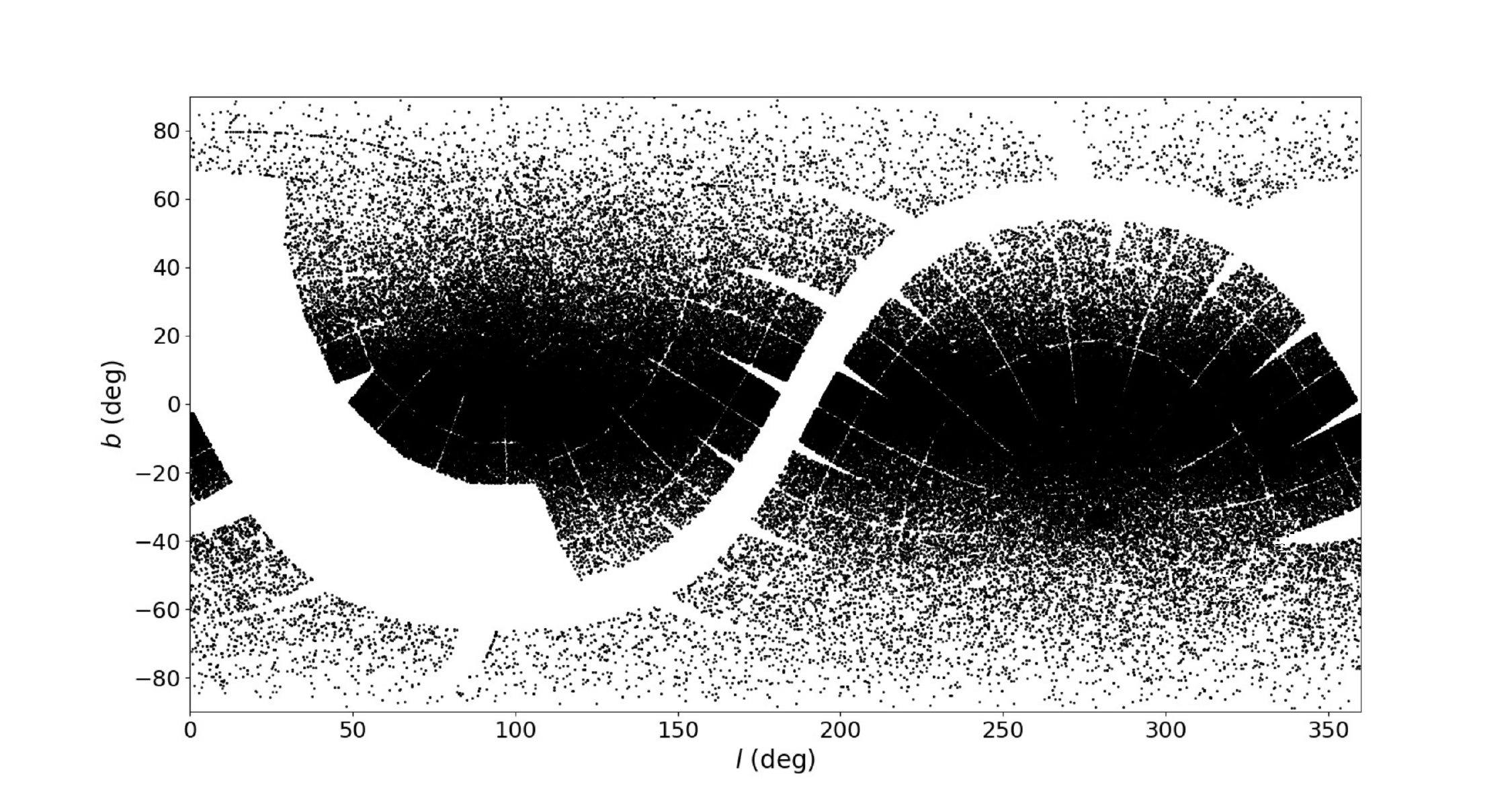}
    \caption{Galactic coordinates of each of the 460,000+ EB candidates found using a neural network on all {\em TESS} light curves up to 15th magnitude in sectors 1-26.}
    \label{fig:ffi_ebcat_coords}
\end{figure}

\end{description}

The catalog of EBs observed by \tess in short cadence is thus only the first step in the exploration of the physical, dynamical and statistical properties of EBs across the entire sky. The catalog will continue to be updated with new sectors of data as they become available, and it will be broadened to include all EB sources detected in the \tess Full-Frame Images. For the latest version of the catalog please refer to \url{http://tessEBs.villanova.edu}.

\begin{acknowledgements}

This paper makes use of data collected by the TESS mission, which are publicly available from the Mikulski Archive for Space Telescopes (MAST). Funding for the TESS mission is provided by NASA’s Science Mission directorate. We acknowledge the use of public \tess data from pipelines at the \tess Science Office and at the \tess Science Processing Operations Center. Resources supporting this work were provided by the NASA High-End Computing (HEC) Program through the NASA Advanced Supercomputing (NAS) Division at Ames Research Center for the production of the SPOC data products.

AP acknowledges support from NASA TESS GI programs G04171, G022062 and G011154. LIJ acknowledges funding from the European Research Council (ERC) under the European Union’s Horizon 2020 research and innovation programme (grant agreement N$^\circ$670519: MAMSIE) and from the Research Foundation Flanders (FWO) under grant agreement 1124321N (Aspirant Fellowship). CJ has received funding from NOVA, the European Research Council under the European Union's Horizon 2020 research and innovation programme (N$^\circ$670519:MAMSIE), and from the Research Foundation Flanders under grant agreement G0A2917N (BlackGEM). KH would like to thank NASA for their continued support through NASA ADAP program 80NSSC19K0594.

This work makes use of Python (Python Software Foundation. Available at \href{http://www.python.org}{www.python.org}) and the Python packages Django \citep{django}, Numpy \citep{numpy}, Numba \citep{numba}, Scipy \citep{scipy} and Matplotlib \citep{matplotlib}.   
\end{acknowledgements}

\bibliographystyle{apj}
\bibliography{references}
\end{document}